\documentclass{JFM-FLM_Au}
\usepackage{graphicx}
\usepackage{epstopdf,epsfig}
\usepackage[utf8]{inputenc}
\usepackage{natbib}
\usepackage{mathrsfs,gensymb}
\usepackage{amsmath,amssymb}
\usepackage{setspace,xcolor}
\usepackage{lscape,comment,multirow,appendix}

\newcommand{\bu}{\mathbf{u}}
\newcommand{\be}{\mathbf{e}}

\newcommand{\pg}[1]{{\color{black}#1}} 

\lefttitle{P. Garaud et al.}
\righttitle{Journal of Fluid Mechanics}

\title{\pg{Two pathways to diapycnal mixing in strongly stratified flows with no initial vertical shear}}

\author{Pascale Garaud\aff{1}, Dante Buhl\aff{1}, Jason Johnstone\aff{1}, Arstanbek Tulekeyev\aff{1}, Nathan Van Duker\aff{1} \corresau{\email{pgaraud@ucsc.edu}}}

\affiliation{
\aff{1} Department of Applied Mathematics, Baskin School of Engineering, University of California Santa
Cruz, Santa Cruz, CA 95064, USA}

\begin{document}

\maketitle

\begin{abstract}
    While vertically-sheared stratified flows have been studied extensively,  their horizontally-sheared counterparts have received considerably less attention. Yet, horizontal shear instabilities remain active even when the \pg{mean} Richardson number is large or even  formally infinite, \pg{and can} drive turbulence \pg{in strongly stratified (low Froude number) flows} at sufficiently high  Reynolds number. 
    In this work, we combine linear theory with direct numerical simulations to  \pg{investigate two pathways to turbulence in low Froude / high Reynolds number horizontally-sheared flow with no initial vertical shear. In the first pathway,  vertical shear emerges directly from vertically-modulated eigenmodes of the primary horizontal shear instability, and becomes unstable to secondary small-scale Kelvin-Helmholtz (KH) instabilities on the buoyancy scale at sufficiently large buoyancy Reynolds number $Re_b$. In the second pathway, a vertically-invariant eigenmode of the primary horizontal shear instability initially dominates, causing the background flow to evolve nonlinearly into a long-lived time-dependent two-dimensional (columnar) vortical flow. The vortices are subsequently unstable to secondary three-dimensional hyperbolic instabilities from which vertical shear emerges, which is finally unstable to tertiary small-scale KH instabilities on the buoyancy scale at sufficiently large $Re_b$.
    This shows that the emergence of vertical shear driving small-scale KH instabilities is an inevitable by-product of horizontal shear instabilities in strongly stratified flows at sufficiently large $Re_b$. However, we also find that the two pathways excite different ranges of vertical scales, which results in  different peak mixing efficiencies.  } 
    \end{abstract}


\section{Introduction}
\label{sec:intro}

The stability of large-scale, strongly stratified horizontal flows to three-dimensional perturbations is a central question for the downscale transfer of energy from the externally forced global scale to the dissipation scales in the Earth's atmosphere and oceans. 
It is also key to the transport of buoyancy, momentum, and passive tracers across diapycnal surfaces. These mechanisms collectively control the momentum and energy balances that regulate the large scale atmospheric and oceanic  circulations \citep{Haynes1991,FerrariWunsch2009}. As such, they must be accurately quantified to improve the fidelity of climate models. 

Theoretical investigations into this problem often focus on the linear and nonlinear stability, and subsequent nonlinear evolution, of highly idealized stably stratified horizontal mean flows ${\bf U}_\star$, with a vertical stratification characterized by a buoyancy frequency $N_{\star}$, see the review by \citet{DrazinReid2004} for examples and references. Although rotation is sometimes taken into account, and likely plays a dominant role at the largest scales on Earth, it is often initially neglected to simplify the problem. 
Even under these restrictive assumptions, a variety of instabilities can arise, depending on the assumed form of the density stratification (e.g. smooth or stepped profiles), the orientation of the shear with respect to the stratification (parallel or inclined), and the flow topology (plane parallel or not). 
These instabilities differ not only in their driving mechanisms but also in their pathways to nonlinear saturation and turbulence, as well as in their mixing properties \citep[see review by][]{Caulfield2021}. It is therefore important to study them systematically, and in isolation, to advance our understanding of their dynamics.

In this paper,  we restrict our analysis to a class of instabilities \pg{that develop in} a background flow with no initial vertical shear  ($\partial {\bf U}_\star/\partial z_\star = 0$, where $z_\star$ is the vertical coordinate), \pg{in a fluid that is linearly stratified in the vertical direction with a constant buoyancy frequency $N_\star$. We specifically focus on the limit of strong stratification, characterized by $Fr \ll 1$}  where $Fr = U_\star/N_\star L_\star$ is the Froude number and $L_\star$ is the characteristic horizontal length scale of the flow ${\bf U}_\star$. Note that here and in all that follows, the star subscript is used to denote dimensional quantities, and any quantity without this subscript is  dimensionless.  \pg{An important question is whether these flows can ultimately generate diapycnal transport and, if so, through what mechanism. Since this requires vertical motions, it must involve the excitation of three-dimensional (3D) perturbations from the initially two-dimensional (2D) flow. In this work, we study two pathways through which this can happen.} 

The simplest possible case within this framework is a plane-parallel flow of the form ${\bf U}_\star = \bar u_\star(y_\star) {\bf e}_x$, the ${\bf e}_x$ direction henceforth always representing the streamwise direction, \pg{and $y_\star$ being the spanwise coordinate}. Various functional forms for $\bar u_\star(y_\star)$ have been explored in past studies, including a hyperbolic tangent profile  \citep{POTYLITSIN_PELTIER_1998,basak2006,Deloncle2007,AroboneSarkar2012,AroboneSarkar2013,LewinCaulfield2024}, a linear profile \citep{Facchinial2018}, \pg{a parabolic profile (Poiseuille flow) \citep{GageReid1968,LeGal2021}} and a sinusoidal profile  \citep{Lucasal2017,cope2020,Garaud2020,Garaudal2024}. 

\pg{We further restrict our analysis to the case where the background flow has an inflection point. In this case,}  the fastest-growing linearly unstable mode is  vertically invariant (i.e. with a vertical wavenumber $k_{z\star} = 0$), driven by a 2D horizontal shear instability \citep[see, e.g.][]{Deloncle2007,AroboneSarkar2012,cope2020}. As the perturbation flow is perpendicular to the density gradient, \pg{this mode is not sensitive to the stratification, but neither does it cause any diapycnal mixing}. Surprisingly, however, the same studies have shown that strong stratification can destabilize 3D ($k_{z\star} \ne 0$) modes that would be stable in more weakly stratified flows. This counterintuitive destabilizing effect of stratification \pg{for $k_{z\star} \neq 0$ modes has recently been explained by \citet{Cocusseetal2025} using asymptotic analysis in the limit of low viscosity and strong stratification. They showed that when a 2D ($k_{z\star}=0$) mode is unstable for some value of the streamwise wavenumber $k_{x\star}$, with growth rate denoted as $\lambda_\star(k_{x\star},0)$, then 3D ($k_{z\star} \neq 0$) modes also exist, and have a growth rate $\lambda_\star(k_{x\star},k_{z\star})$ that asymptotically tends to $\lambda_\star(k_{x\star},0)$ when $Fr \rightarrow 0$. These eigenmodes become increasingly anisotropic with increasing stratification, with  vertical velocities scaling as $Fr^2$ times their horizontal velocities, as in the Lilly regime \citep{Lilly1983}.
 
 Direct Numerical Simulations (DNS) are required to study the evolution of an instability beyond linear theory and compute its diapycnal mixing properties. This can be done either as an initial value problem (IVP), or as a forced problem. In the IVP, numerical studies of a stratified hyperbolic tangent shear layer} have found that  $k_{z\star} = 0$ modes of instability first grow \pg{to dominate the flow} and nonlinearly evolve into \pg{a row of} vertically-invariant `columnar'  vortices, at least initially \citep{POTYLITSIN_PELTIER_1998,basak2006,AroboneSarkar2013}. \pg{The same studies found that 3D perturbations with vertical wavenumber $k_{z\star} L_\star = O ( Fr^{-1})$} also emerge albeit later, and grow provided the Reynolds number $Re = U_\star L_\star/\nu_\star$ (where $\nu_\star$ is the kinematic viscosity of the fluid) is large enough. \pg{This secondary instability of the columnar vortices has been identified as the so-called `zigzag' instability \citep{BillantChomaz2000_linear,billant2001}, which takes place in neighboring vortex pairs. \citet{deloncle2008} showed that the growth of the zigzag instability is either halted by viscous effects when the buoyancy Reynolds number $Re_b = Fr^2 Re =  O(1)$, or by tertiary small-scale Kelvin-Helmholtz (KH) instabilities of the emergent vertical shear for sufficiently large $Re_b$ \citep[see also the more recent works of][on the nonlinear development of the zigzag instability in a Taylor-Green vortex array]{Guoal2024,Guoal2025}.}

\begin{figure}
\includegraphics[width=\textwidth]{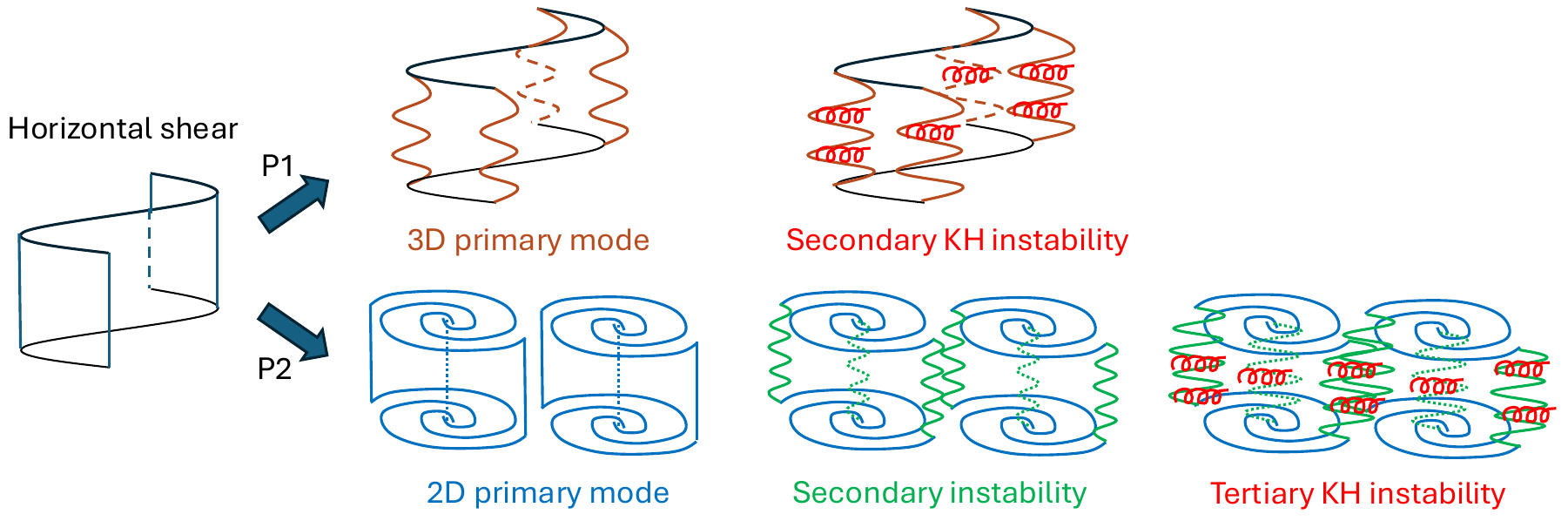}
\caption{\pg{Illustration of the two possible pathways from a vertically-invariant horizontal shear flow, to the development of small-scale KH instabilities of the emergent vertical shear. This assumes that $Fr \ll 1$ (to ensure that 3D primary modes are unstable) and $Re_b \gg 1$ (to ensure that KH instabilities on small scales can develop). The top row illustrates the pathway P1  described for instance in the body-forced DNS of a sinusoidal flow by  \citet{cope2020} and \citet{Garaud2020}, while the bottom row illustrates the pathway P2 described in the IVP DNS of a hyperbolic tangent profile by \citet{basak2006} and \citet{deloncle2008}. }}
\label{fig:illustration}
\end{figure}

\pg{By contrast with the IVP where the flow inevitably decays, a statistically stationary turbulent state can always be reached in the body-forced problem (at sufficiently high Reynolds number). Focussing on prior works that used a vertically-invariant body force only, two approaches have been proposed: time-dependent stochastic forcing \citep[see e.g.][]{WAITE_BARTELLO_2004,brethouwer2007,augier2015,maffioli2016}, and steady forcing  \citep{Lucasal2017,cope2020,Garaud2020,Garaudal2024}. Curiously, the emergence of 3D structures in simulations with steady forcing appears to be rather different from the sequence of instabilities that take place in the initial value problem described above. In DNS that use a sinusoidal body force (thus driving what is commonly known as a Kolmogorov flow), $k_{z\star} \neq 0$ perturbations were observed to dominate even at early times \citep{cope2020}, thus bypassing the aforementioned columnar flow stage. These generate vertical shear on a broad range of scales \citep[consistent with][]{Cocusseetal2025}, that later becomes unstable to secondary small-scale KH instabilities provided  $Re_b$ is sufficiently large. 

These prior studies therefore 
show the existence of  two  distinct pathways (illustrated in figure \ref{fig:illustration}) from a 2D plane-parallel strongly stratified horizontally-sheared flow to the development of 3D perturbations with vertical shear, followed by small-scale vertical KH instabilities. This raises the following questions. (1) 
What determines which pathway is selected by the flow? Is it the inherent difference between an IVP vs. a  body-forced problem, or is it due to other differences between the aforementioned studies, such as the shape of the initial flow (hyperbolic tangent vs. sinusoidal, selection of the specific form of the initial conditions), or the boundary conditions applied (wall-bounded vs. periodic)? For example, would an initial value problem initialized with a sinusoidal profile instead of a hyperbolic tangent profile follow pathway P1 as in \citet{cope2020} (first through the development of vertical shear through the growth of $k_z \neq 0$ modes, and then via KH instabilities of that emergent vertical shear), or would it follow pathway P2 as in \citet{basak2006} (first through the development of columnar vortices, followed by a zigzag instability and tertiary KH instabilities)? And (2), does the pathway selected have any impact on  the peak diapycnal mixing and mixing efficiency during the nonlinear development of the instability?   }

\pg{To answer question (1), we begin in  \S\ref{sec:DNS} by studying the nonlinear development of a horizontal, plane-parallel, strongly-stratified sinusoidal flow in a triply-periodic domain, in the IVP. We demonstrate that both pathways to the small-scale 3D KH instabilities can be achieved in exactly the same model setup simply by changing the initial conditions slightly to favor the $k_{z\star} = 0$ mode, or not. In \S\ref{sec:secondary}, we study in more detail the nature of the secondary instability of the columnar flow that develops in the second pathway in our simulations. We show that it is a hyperbolic instability that shares strong similarities with the zigzag instability, and notably develops on very small vertical length scales. In \S\ref{sec:eta}, we study diapycnal mixing and energy dissipation in both pathways to answer question (2).}  
  We summarize our results in \S\ref{sec:ccl}, and outline  questions that remain to be answered to apply this work in the context of geophysical and astrophysical flows.

\section{\pg{Nonlinear evolution of strongly stratified Kolmogorov flows: two possible pathways}}
\label{sec:DNS}

We therefore 
consider the vertically invariant, horizontal, plane-parallel Kolmogorov flow
\begin{equation}
    \bar \bu_\star(y_\star) = U_\star \sin(k_\star y_\star) \be_x, 
\end{equation}
in an incompressible stratified fluid with constant buoyancy frequency $N_\star$, where the vertical direction ${\bf e}_z$ is (anti-)aligned with gravity.
We set the inverse wavenumber $k_\star^{-1}$ and the amplitude $U_\star$ of this flow to be the unit  length and  velocity, respectively, so the corresponding dimensionless Kolmogorov flow is
\begin{equation}
\bar \bu(y) = \sin(y) \be_x. 
\label{eq:ubardef}
\end{equation}

\pg{In this section, we use DNS to study the nonlinear evolution of $\bar \bu$ subject to small-amplitude initial perturbations. 
This evolution is governed by 
the} incompressible  dimensionless Navier-Stokes equations in the Boussinesq approximation: 
\begin{eqnarray}
    \frac{\partial  \bu}{\partial t} + \bu \cdot \nabla \bu  = - \nabla  p + \frac{ b}{Fr^2} \be_z +  \frac{1}{Re} \nabla^2 \bu,  \label{eq:DNSumom}\\
\frac{\partial b}{\partial t} + \bu \cdot \nabla  b + w =   \frac{1}{Pe} \nabla^2  b, \\ 
    \nabla \cdot \bu = 0,  \label{eq:DNSudiv}
\end{eqnarray}
\pg{where ${\bf u} = (u,v,w)$ is the flow field in units of $U_\star$, $p$ is the pressure perturbation away from hydrostatic equilibrium in units of $\rho_\star U_\star^2$ (where $\rho_\star$ is the mean density of the fluid),
and $b$ is the buoyancy perturbation away from the linearly stratified background expressed in units of $N_\star^2/k_\star$. Lengths and time are in units of $L_\star$ and $L_\star/U_\star$, respectively.} All dependent quantities (${\bf u}, p, b$) are assumed to be triply-periodic in the domain. 
 The dimensionless parameters that appear are the usual Reynolds, P\'eclet and Froude numbers:
\begin{equation}
Re = \frac{U_\star}{k_\star \nu_\star}, \quad Pe = \frac{U_\star}{k_\star \kappa_\star}, \quad Fr = \frac{U_\star k_\star}{N_\star}, 
\end{equation}
where $\nu_\star$ is the kinematic viscosity and $\kappa_\star$ is the buoyancy diffusivity. 

\pg{To focus on the limit of strong stratification,  we explore two different values of the Froude number: $Fr = 0.15$ and $Fr = 0.1$. As discussed by others before \citep[see the review by][]{Caulfield2021},  turbulence in low $Fr$ flows can only be achieved at sufficiently large buoyancy Reynolds number. For this reason, we further select $Re = Pe = 10000$, so $Re_b = Fr^2 Re = 225$ when $Fr = 0.15$, and $Re_b = 100$ when $Fr = 0.1$}.
 
\pg{At these parameters, a substantial resolution is needed to resolve all scales. For this reason, we are forced to use a relatively small computational domain, of size $4\pi \times 2\pi \times \pi$. The selected width $L_y = 2\pi$ is exactly one wavelength of the background Kolmogorov flow, which forces perturbations to have the same periodicity. The length $L_x = 4\pi$ is chosen to be close to (but slightly larger than) the wavelength of the fastest-growing mode of the horizontal shear instability for this flow \citep[see, e.g.][]{cope2020}. Finally the depth $L_z = \pi$ selected is large enough to contain many buoyancy length scales $L_b = Fr$ for both values of $Fr$. This scale is the expected scale of the zigzag instability \citep{BillantChomaz2000_theory}, and of turbulent eddies in strongly stratified turbulence \citep{brethouwer2007,Riley_Lindborg_2013,chini2022}. We note that while larger domain sizes in both horizontal directions would allow for an even richer range of dynamics, the choice made here is sufficient for the purpose of this work, namely to compare the two pathways P1 and P2  to three-dimensionality and stratified turbulence in strongly stratified horizontally-sheared flows. }

\begin{figure}
\begin{center} \includegraphics[width=\textwidth]{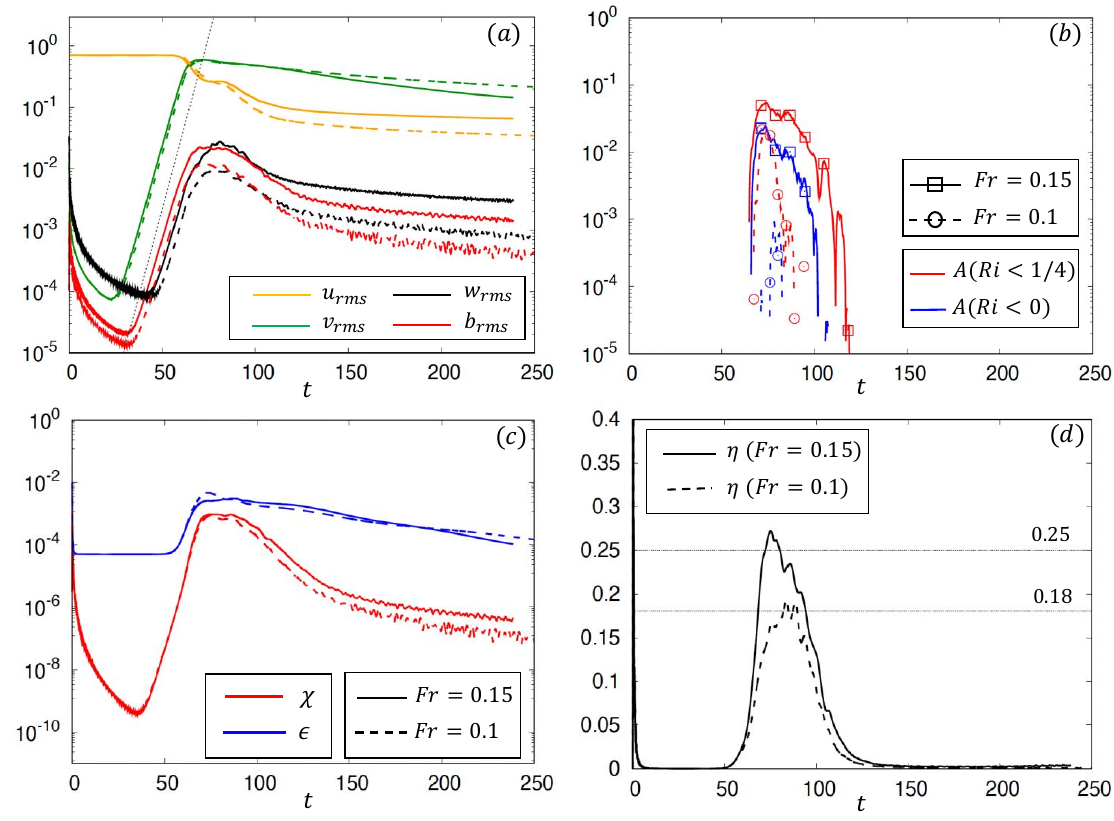}
\caption{\pg{Results from  the set of DNS1  described in \S\ref{subsec:DNS1}, with $Re = Pe = 10000$, $Fr = 0.15$ (solid lines, DNS1a) and $Fr = 0.1$ (dashed lines, DNS1b). ({\it a}) Rms velocities  $u_{\rm rms}$ (orange), $v_{\rm rms}$ (green) and $w_{\rm rms}$ (black), and buoyancy fluctuation $b_{\rm rms}$ (red). The black dotted line shows the predicted growth of the fastest-growing mode of primary instability (with growth rate $\lambda = 0.26$). ({\it b}) Area fraction of the $y=0$ plane (lines) and volume fraction of the domain (symbols) occupied by regions where $Ri < 1/4$ (red) and $Ri < 0$ (blue).     
    ({\it c}) Scaled dissipation of buoyancy variance $\chi$ and viscous dissipation of kinetic energy  $\epsilon$  (see equation \ref{eq:epschidef}). ({\it d}) Mixing efficiency $\eta$ (defined in equation \ref{eq:etadef}).}}
\label{fig:DNS1}
\end{center}
\end{figure}

\begin{figure}
\begin{center} \includegraphics[width=\textwidth]{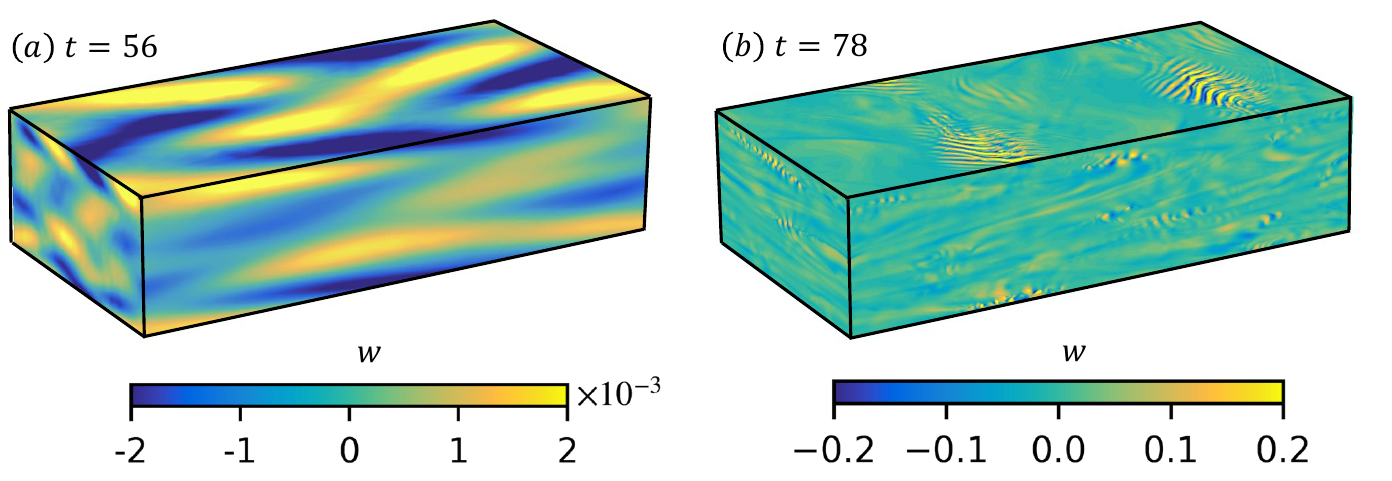}
\caption{\pg{Snapshots of the vertical velocity field $w$ for DNS1a ($Re = Pe = 10000, Fr = 0.15$), initialized with white noise. ({\it a}) During the exponential
growth phase of the primary mode of instability at $t = 56$. ({\it b})  During the saturation of the primary mode
of instability at $t = 78$. } }
\label{fig:DNS1_snaps}
\end{center}
\end{figure}

The simulations are performed using the pseudo-spectral PADDI code \citep{Traxleral2011}, with a resolution of $1536 \times 768 \times 384$ equivalent grid points. \pg{This corresponds to a grid scale that is equal to approximately twice the Kolmogorov scale at the time of peak kinetic energy dissipation for all simulations.  

 The buoyancy perturbations are initially set to zero, namely 
\begin{equation}
    b(x,y,z,0) = 0, 
    \label{eq:initcondb}
\end{equation}
while the flow is initialized as
\begin{equation}
    \bu(x,y,z,0) = \bar \bu(y) + \tilde{\bu}_0(x,y,z),
    \label{eq:initcondu}
\end{equation}
where different initial perturbations $\tilde{\bu}_0$ are explored in the following sections.

\subsection{\pg{Initial grid-scale noise (DNS1) } }
\label{subsec:DNS1}

 We begin with a first set of DNS (DNS1 hereafter) initialized with grid-scale noise 
\begin{equation}
    \tilde \bu_0 (x,y,z) = {\bf n}(x,y,z),
\label{eq:initcondDNS1}
\end{equation}
generated as follows.  We first  create random white noise on the grid scale and assign it to the vertical component of the velocity field $\tilde \bu_0$.} A divergence-cleaning step is then applied to ensure that $\tilde \bu_0$ is divergence-free at $t= 0$, which has the effect of adding some noise to the other two components of $\tilde\bu_0$. 
\pg{Finally, the characteristic amplitude of the noise is scaled to be $O(0.01)$.}
This type of initial condition was used in several prior investigations of forced Kolmogorov flows \citep{cope2020,Garaud2020,Garaudal2024}. \pg{We note that viscous diffusion and buoyancy diffusion rapidly damp out fluctuations on the grid scale, resulting in much smoother initial conditions for all practical purposes.

Based on the linear stability analysis of \citet{Cocusseetal2025}, we anticipate the development of perturbations on large horizontal scales, with a broad range of vertical wavenumbers $k_z \ll Fr^{-1}$  growing exponentially at a rate close to that of the fastest-growing 2D mode that fits in the computational domain. As summarized in Appendix A, this mode has a horizontal wavenumber $k_x = 0.5$ (when forced to fit in a domain of length $L_x = 4\pi$) and a growth rate $\lambda \simeq 0.26$. We also anticipate that the perturbation field should be highly anisotropic, with characteristic vertical velocities that are smaller than the horizontal velocities by a factor $O(k_z^2 Fr^2)$. 

  Figure \ref{fig:DNS1} shows the temporal evolution of several quantities of interest for two DNS at $Re = Pe = 10000$, $Fr = 0.15$ (DNS1a) and $Fr = 0.1$ (DNS1b), respectively. Figure \ref{fig:DNS1_snaps} shows snapshots of the vertical velocity field for DNS1a during the linear growth phase of the primary instability ($t = 56$), and around  saturation ($t = 78$). Selected fields ($u$ and $w$) are shown in movie 1, also for DNS1a. Overall, the evolution of the system is consistent with the pathway P1 described in figure \ref{fig:illustration}.}
\\
\\
\noindent \pg{{\sc  Movie  1}: Evolution of the streamwise velocity $u$ (left) and of the vertical velocity $w$ (right) throughout the domain, in DNS1a ($Re = Pe = 10000$, $Fr = 0.15$), initialized with white noise}.
\\
\\

\pg{At early times ($t < 60$), the results are consistent with expectations from linear theory. }  Movie \pg{1} shows that three-dimensional structures \pg{with large horizontal scales and a wide range of vertical scales} emerge directly from the initial conditions. Figure
 \ref{fig:DNS1}({\it a}) shows the temporal evolution of the rms velocities $u_{\rm rms} = \langle u^2 \rangle^{1/2}$, $v_{\rm rms} = \langle v^2 \rangle^{1/2}$ and $w_{\rm rms} = \langle w^2 \rangle^{1/2}$, where the angular brackets denote a volume average. It reveals that both $v_{\rm rms}$ and $w_{\rm rms}$ grow exponentially at the same rate, \pg{which is close to and slightly smaller than that of the 2D mode $\lambda = 0.26$ (see black dotted line). Finally, we see that $w_{\rm rms} \sim b_{\rm rms} \simeq Fr^2 v_{\rm rms}$, as predicted by the asymptotic analysis of \citet{Cocusseetal2025}. Taken together, these clues  clearly identify the $k_z \ne 0$ modes of instability of the initial flow $\bar {\bu}$ as the origin of these 3D structures.} 

 A visual inspection of the horizontal velocity field (left panel of movie \pg{1}) around $t = 56$ reveals a chevron pattern in the $x=0$ plane, that is now well-known in the context of stratified Kolmogorov flows \citep{Lucasal2017,cope2020}.  We see that this pattern simply corresponds to meandering perturbations of the mean flow (see the $z = \pi$ plane) whose phases vary in the vertical direction. 

The growth of the primary $k_z \ne 0$ perturbations saturates when $v_{\rm rms} \simeq u_{\rm rms}$, around $t = 60$ (see figure \ref{fig:DNS1_snaps}({\it a})). However, $w_{\rm rms}$ continues to grow beyond that point until $t \simeq 80$, albeit more slowly. Inspection of the vertical velocity field at $t = 78$ (see figure \ref{fig:DNS1_snaps}({\it b})) shows the signature of small-scale, localized KH instabilities. \pg{These are caused by the strong vertical shear that emerges from the $k_z\neq 0$ modes of the horizontal shear instability (see movie  1), and appear as ripples on the top face of the domain, and regions of strong $|w|$ on the faces of the domain}.    

\pg{In a steady plane parallel flow, KH instabilities can be triggered when the  Richardson number
drops below 1/4 somewhere in the domain \citep{Miles1961,Howard1961}. While this criterion does not directly apply here  because the emergent vertical shear is neither steady nor plane parallel, it is also often argued \citep[see, e.g.][]{Richardson1920} that turbulent eddies can gain energy in a stratified shear flow wherever the local Richardson number $Ri$ is small enough, where we   
 define $Ri$ here as} 
\begin{equation}
Ri = \frac{1}{Fr^2} \frac{1 + \frac{\partial b}{\partial z} } {\left( \frac{\partial u}{\partial z} \right)^{2} + \left( \frac{\partial v}{\partial z} \right)^{2}}.  
\end{equation} 
\pg{ Figure \ref{fig:DNS1}({\it b})} shows the fractional area of the $y = 0$ plane \pg{(lines) and fractional volume of the whole domain (symbols)} occupied by regions where $Ri < 1/4$, in \pg{DNS1a ($Fr = 0.15$, solid red line) and in DNS1b ($Fr = 0.1$, dashed red line). The volume fraction data is only available at times when full simulation data was saved, while the area fraction data is available at much more frequent intervals. The comparison between the two shows that the area fraction reliably reproduces the volume fraction despite probing only a single plane.  We see that starting around $t = 60$ a small, but non-negligible fraction of the domain satisfies this local Richardson criterion in both cases. Not surprisingly, we see that this effect is more pronounced and lasts for longer in the more weakly stratified case ($Fr = 0.15)$. Figure \ref{fig:DNS1}({\it b})} similarly shows in blue the fractional \pg{area/volume} of the domain that is occupied by regions where $Ri <0$ in the same  DNS. These are regions where a KH billow has caused a local inversion in the density gradient. We see that both \pg{DNS1a and DNS1b} contain such density inversions (solid and dashed lines), \pg{and that their volume fraction is largest between $t = 70$ and $t = 90$, which corresponds to the time in figure \ref{fig:DNS1}({\it a}) where $w_{\rm rms}$ peaks. As expected, the more strongly stratified case has much less frequent overturns, resulting in a smaller peak $w_{\rm rms}$.}

\pg{Finally, we} define the instantaneous mixing efficiency as
\begin{equation}
    \eta(t) = \frac{ \chi(t) }{\chi(t)  + \epsilon(t)},
\label{eq:etadef}
\end{equation}
\pg{where the scaled buoyancy dissipation rate $\chi$ and the viscous disspation rate of kinetic energy $\epsilon$ are defined as} 
 \begin{equation}
  \chi(t) = \frac{1}{Fr^2Pe} \langle|\nabla b|^2\rangle(t) , \quad \epsilon(t) = \frac{1}{Re} \langle | \nabla {\bf u} |^2 \rangle(t).
  \label{eq:epschidef}
\end{equation}  
 
 We see in figure \pg{\ref{fig:DNS1}({\it c,d}) that $\chi$, $\epsilon$ and $\eta$ all} peak between $t = 70$ and $t =90$, \pg{i.e. the time with highest KH activity.  The maximum value
$\eta_{\rm max} \simeq 0.18$ for $Fr = 0.1$, and $\eta_{\rm max} \simeq 0.25$ for $Fr = 0.15$.} 

\subsection{\pg{Initial grid-scale noise + small amplitude 2D mode (DNS2)}}
\label{subsec:DNS2}

The \pg{second set of} DNS presented (DNS\pg{2} hereafter) is initialized with  \pg{grid-scale noise {\bf n} generated as described in the previous section, plus a `seed' for the $k_z = 0$ mode of the primary instability, denoted  as $\bu_{\rm s}(x,y)$. These  initial conditions are given by equations  (\ref{eq:initcondb}) and (\ref{eq:initcondu}) where this time
\begin{equation}
    \tilde\bu_0(x,y,z,0) = \bu_{\rm s}(x,y) + {\bf n}(x,y,z).
    \label{eq:initcondDNS}
\end{equation}
The seed $\bu_{\rm s}(x,y)$ is chosen to closely approximate the fastest-growing 2D eigenmode of the horizontal shear instability  of the background flow $\bar \bu$ (which has $k_z = 0$ and $k_x =0.5$, see Appendix A for detail), with amplitude $a = 0.1$:
\begin{equation}
    \bu_{\rm s}(x,y) = a \tilde \bu_{\rm m}(x,y) = a \left[ u_{\rm m} (x,y) \be_x + v_{\rm m}(x,y) \be_y 
\right] ,
\end{equation}
}
where the subscript m stands for `meander', and 
\begin{equation}
   u_{\rm m}(x,y) = 2 \cos(y) \cos(k_x x), \quad v_{\rm m}(x,y) = v_0 \cos(k_x x) + \sin(y) \sin(k_x x), 
   \label{eq:meanderdef}
\end{equation}
with $v_0 \simeq -0.913$. We show in Appendix A that $\tilde{\bf u}_{\rm m}(x,y)$ is quite close to the exact \pg{linear eigenmode} $\tilde{\bf u}(x,y)$ for $k_x = 0.5$, $k_z = 0$. 
\\
\\
\noindent {\pg{{\sc Movie 2}: Evolution of the vertical vorticity at $z = 0$ (left) and of the vertical velocity throughout the domain (right), in DNS2a  with $Re = Pe = 10000$, \pg{$Fr = 0.15$}, and initial conditions given in equation \eqref{eq:initcondDNS}.} 
\\
\\
\pg{The temporal evolution of quantities of interest are shown in figure \ref{fig:DNS2} for two simulations at $Fr = 0.15$ (DNS2a, solid lines) and $Fr = 0.1$ (DNS2b, dashed lines), respectively. Snapshots of $w$ are shown in figure \ref{fig:DNS2_snaps} for the $Fr = 0.15$ case}. 
Movie \pg{2} shows the temporal evolution of the vertical vorticity field $\omega_z = (\nabla \times {\bf u})\cdot {\bf e}_z$ on the $z = 0$ plane (left) and of the vertical velocity field $w$ in the whole domain (right), \pg{in DNS2a. It} illustrate the rich dynamics of this model system from the initial growth of the meander\pg{ing perturbation} to the ultimate decay of the stratified turbulence generated by successive instabilities \pg{consistent with the pathway P2 described in figure \ref{fig:illustration}}. 

At early times, movie \pg{2} shows a rapid growth in the amplitude of the seeded meander, driven by the primary horizontal shear instability,  see  Appendix A. Correspondingly,  
$v_{\rm rms}$ increases exponentially with a growth rate approximately equal to $\lambda = 0.26$ (shown in the dashed green line in figure \ref{fig:DNS2}({\it a})), which is consistent with the predicted growth rate for the $k_x = 0.5, k_z = 0$ mode. At the same time,  $w_{\rm rms}$ decays rapidly with time due to viscosity damping out the  grid-scale noise. 

The primary instability saturates nonlinearly around $t=10$, which causes a rapid decrease in the streamwise flow amplitude ($u_{\rm rms}$). 
Crucially, \pg{and in striking contrast with DNS1, we see that} the flow then remains primarily 2D (columnar) until $t = 120$, and
takes the form of roughly circular vortices with a radius close to one, that are advected in the $y$ direction by a more-or-less sinusoidal flow $\propto \sin (k_x x) {\bf e}_y$ (see movie \pg{2}). \pg{We conclude that the $k_z \neq 0$ modes of instability of $\bar \bu$ discussed by \citet{Cocusseetal2025} are unable to grow when the background flow is substantially affected by the presence of a dominant $k_z = 0$ mode. Why this is the case remains an open question.}

\begin{figure}
\includegraphics[width=\textwidth]{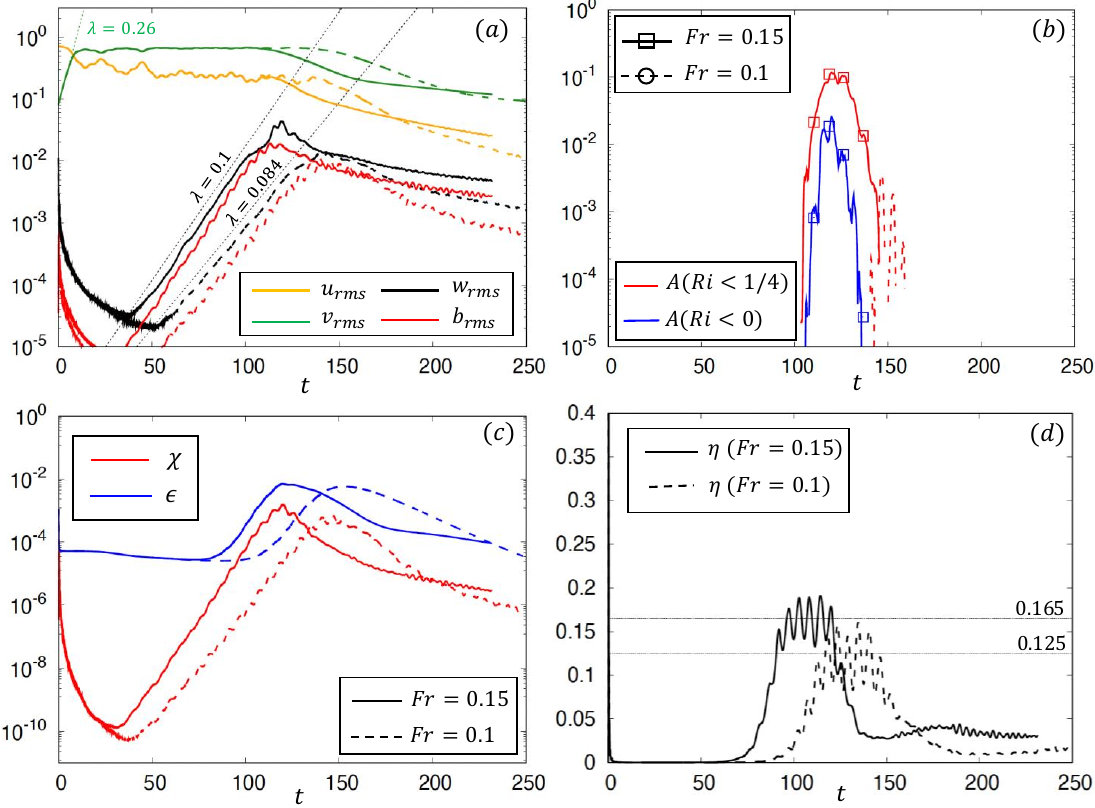}
    \caption{\pg{Results from the set of DNS2 described in \S\ref{subsec:DNS2}, with $Re = Pe = 10000$, $Fr = 0.15$ (solid lines, DNS2a) and $Fr = 0.1$ (dashed lines, DNS2b). ({\it a}) Rms velocities  $u_{\rm rms}$ (orange), $v_{\rm rms}$ (green) and $w_{\rm rms}$ (black), and buoyancy fluctuation $b_{\rm rms}$ (red). The green dashed line shows the predicted growth of the corresponding fastest-growing mode of primary instability (with growth rate $\lambda = 0.26$), and the black dotted lines are fits to the observed exponential growth of the secondary mode of instability, with  estimated growth rates listed in the figure.  ({\it b}) Area fraction of the $y=0$ plane (lines) and volume fraction of the domain (symbols) occupied by regions where $Ri < 1/4$ (red) and $Ri < 0$ (blue).     
    ({\it c}) Scaled dissipation of buoyancy variance $\chi$ and viscous dissipation of kinetic energy  $\epsilon$ in the same simulations (see equation \ref{eq:epschidef}). ({\it d}) Mixing efficiency $\eta$ (defined in equation \ref{eq:etadef}). }} 
\label{fig:DNS2}
\end{figure}

\begin{figure}
\begin{center} \includegraphics[width=\textwidth]{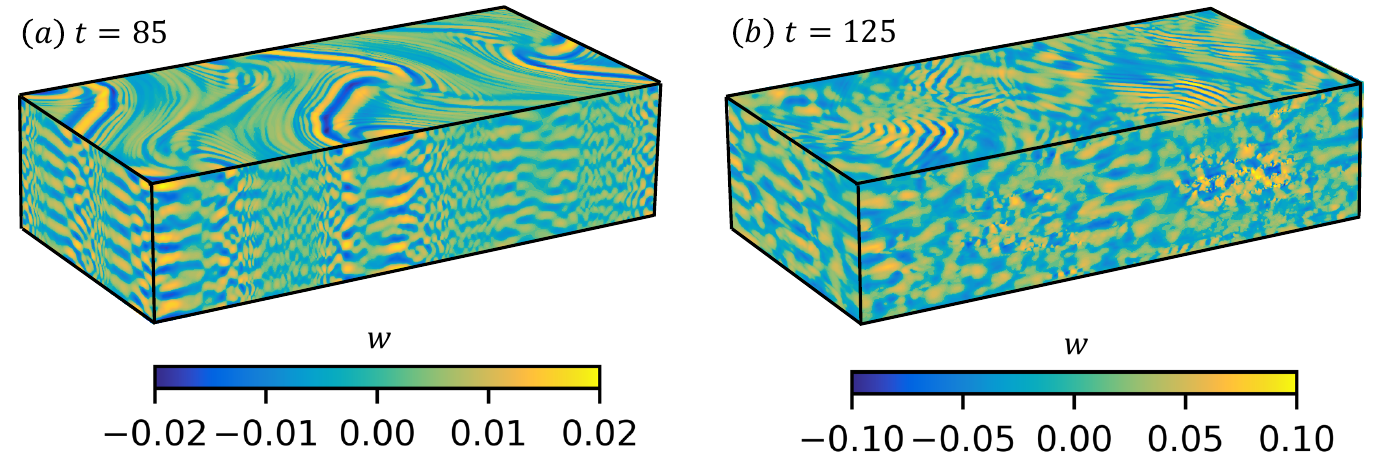}
\caption{\pg{Snapshots of the vertical velocity field $w$ for DNS2a ($Re = Pe = 10000, Fr = 0.15$), initialized with a seed meander for the $k_z = 0$ mode. ({\it a}) During the exponential growth phase of the secondary mode of instability at $t = 85$. ({\it b}) During the saturation of the secondary mode of instability at $t = 125$.}}
\label{fig:DNS2_snaps}
\end{center}
\end{figure}

Starting around $t = 40$, we begin to see signatures of a secondary mode of instability that drives the exponential growth of vertical motions (see figure \ref{fig:DNS2}({\it a})) and of the scaled buoyancy dissipation 
(see figure \ref{fig:DNS2}({\it c})). Fitting an exponential to $w_{\rm rms}(t)$ reveals that the mode growth rate is \pg{approximately equal to $\lambda_{{\rm DNS2a}} \simeq 0.1$  for the $Fr = 0.15$ simulation, and $\lambda_{{\rm DNS2b}} \simeq 0.084$  for the $Fr = 0.1$ simulation}. 
The evolving vertical velocity  (see movie \pg{2} and figure \ref{fig:DNS2_snaps}({\it a})) shows what appears to be ripples with a short horizontal length scale. Inspection of the vertical velocity field on the $y=0$ plane at the same time shows that this mode has a vertical wavenumber \pg{$k_z = 12$ (i.e. it has 6 periods in a domain of height $L_z = \pi$) for the $Fr = 0.15$ case, and $k_z = 16$ for the $Fr = 0.1$ case (see figure \ref{fig:Datavstheory}({\it d}))}.  Movie \pg{2} shows that these vertical fluctuations are advected in the $y$ direction by the mean flow in conjunction with the vortices.   

The rms vertical velocity $w_{\rm rms}$ peaks when the secondary mode of instability saturates. The viscous dissipation of kinetic energy $\epsilon$
 is about 100 times larger at this point than in the columnar flow phase, see figure \ref{fig:DNS2}({\it c}). This drives a rapid decay of the vortices (see movie \pg{2}), followed by a more gradual decay of the overall kinetic energy of the mean sinusoidal flow $\propto \sin(k_x x){\bf e}_y$ at late times.

\pg{As in \citet{deloncle2008}, we find that the secondary instability saturates in different ways depending on $Re_b$. For the somewhat more weakly stratified case (DNS2a, $Re_b = 225$), we see again the clear signature of tertiary KH instabilities developing on top of the secondary $k_z \neq 0$ modes of instability, both in the snapshot in figure \ref{fig:DNS2_snaps}({\it b}), and in the area fraction occupied by $Ri<0$ regions in figure \ref{fig:DNS2}({\it b}). By contrast these KH instabilities are effectively absent in the more strongly stratified case (DNS2b, $Re_b = 100$), suggesting that the secondary instability saturates viscously in this simulation. Correspondingly, the peak values of $\eta$ close to the saturation of the secondary modes is significantly lower in DNS2b than in DNS2a (see figure \ref{fig:DNS2}({\it d})). Generally, we also find that the maximum of $\eta$ is lower in DNS2 than in DNS1 at the same values of $Fr$, even though the maximum value of $\chi$ is approximately the same. This will be discussed in \S\ref{sec:eta}. }

\pg{

\subsection{Summary of the DNS results}

Through these DNS, we have answered the first set of questions raised in \S\ref{sec:intro}: the two pathways illustrated in figure \ref{fig:illustration}
from an initially vertically-invariant, strongly stratified horizontal shear flow to 
the onset of vertical KH instabilities can both be found in exactly the same model setup, with only small changes in initial conditions, by either exciting all modes of the primary horizontal shear instability equally, or by favoring the $k_z = 0$ mode. 
For perturbations seeded with grid-scale white noise only, $k_z \neq 0$ eigenmodes of the primary horizontal shear instability develop, generating vertical shear that eventually becomes strong enough to trigger secondary KH instabilities, as in the pathway P1 \citep{cope2020}. For initial conditions that  preferentially excite the $k_z = 0$ mode of the primary instability, by contrast, the flow first evolves into a 2D nonlinear columnar state, which is unstable to secondary 3D modes. These have a high $k_z$ and a distinctive horizontal structure, and saturate through tertiary KH instabilities for large enough $Re_b$, as in the pathway P2  \citep{basak2006,deloncle2008}. 

We note that while these results were obtained specifically for the Kolmogorov flow, they should also hold for other horizontal shear flows that have an inflection point (so the primary instability has both $k_z = 0$ and $k_z \neq 0$ modes), such as the commonly used hyperbolic tangent profile. We believe that the main reason why pathway P1 (via $k_z \neq 0$ modes of the primary instability) has not been discussed as much in this context is that DNS of the hyperbolic tangent shear layer are often initialized with a seed for the $k_z = 0$ mode, but this choice prevents P1 from occurring.   

\section{Secondary instability of the columnar flow}
\label{sec:secondary}

In this section, we aim to identify the nature of the high $k_z$ secondary instabilities that develop in DNS2. Indeed, the vertical velocity eigenfunctions shown in figure \ref{fig:DNS2_snaps}({\it a}) and movie 2 do not look like the traditional zigzag instabilities described by \citet{BillantChomaz2000_experiment,BillantChomaz2000_theory}. This is perhaps not surprising in as much as the vortical flow field that develops in these simulations (see movie 2 at early times) looks more like a vortex array than isolated vortex pairs. }

The study of instabilities of 
stratified columnar vortices and vortex arrays dates back to the 1990s with the foundational studies of \citet{MiyazakiFukumoto1992} and \citet{POTYLITSIN_PELTIER_1998}. Isolated columnar vortices, as well vortex pairs or vortex arrays \pg{can be destabilized in many ways} \citep[see the review by][]{Chomazal2010}. For instance, circular isolated vortices are prone to shear and centrifugal instabilities \citep{Rayleigh1892instability,Rayleigh1917}, while strained vortices are subject to the elliptical instability \citep{MooreSaffman1975}, see the review by \citet{Kerswell2002AnRFM}. \pg{The aforementioned} zigzag instability affects both counter-rotating \citep{BillantChomaz2000_linear} and co-rotating \citep{Otheguyal2007} vortex pairs. In vortex arrays, finally, the presence of hyperbolic stagnation points introduces the possibility of hyperbolic instabilities  \citep{FriedlanderVishik1991,Leblanc1997,KerswellCaulfield2000}, that persist in the presence of stratification \citep{Suzukial2018,Hattorial2021,Guoal2024}.

For many of these vortex instabilities, the effect of stratification is subtle: it suppresses certain modes while enhancing others that have a resonance between the vortex circulation and a gravity mode's frequency. Stratification thus strongly affects mode selection, favoring perturbations with a vertical length scale of the order of the buoyancy scale $L_{b} = Fr$  \citep{BillantChomaz2000_theory,Hattorial2021,Guoal2024}. In strongly stratified flows ($Fr \ll 1$) these 3D perturbations are highly anisotropic, \pg{as we find here. We now look at the stability of the columnar flow in our own DNS in more detail, and attempt to identify which of the aforementioned instabilities takes place.}

\subsection{A simplified model of the columnar flow}
\label{sec:ubmodel}

\pg{In order to study the secondary instabilities of the columnar flow that develop in DNS2a and DNS2b, we invoke the frozen-in approximation, which assumes that $k_z \neq 0$ perturbations grow much faster than the rate at which the columnar flow evolves. While this is not necessarily true here, as we shall see, this assumption still provide very useful insight into the nature of these modes. }

We first create a simplified  analytical model of the \pg{columnar flow, which we henceforth call ${\bf u}_{\rm b}$ in reference to the fact that it becomes the new `background' flow. We define it as 
\begin{equation} 
\bu_{\rm b}(x,y) =  \bar \bu(y) + a  \left[ u_{\rm m}(x,y) \be_x + v_{\rm m}(x,y) \be_y 
\right],
\label{eq:ub}
\end{equation} 
where $u_{\rm m}$ and $v_{\rm m}$ are given in equation (\ref{eq:meanderdef}). 
This simply assumes that ${\bf u}_{\rm b}$ is equal to the original background flow ${\bar \bu}$, plus an approximate expression for the $k_z = 0$ eigenmode of the primary horizontal shear instability, $\bu_{\rm m} = (u_{\rm m},v_{\rm m})$,  with some prescribed amplitude $a$.} While $a$ is a constant in this expression, increasing $a$ effectively models the growth of the columnar meander through the primary instability.

Because $\nabla \cdot \bu_{\rm b} = 0$, we can define the  stream function 
\begin{equation}
\psi_{\rm b}(x,y) = \cos(y) - 2a\sin(y)\cos(k_x x) + 2av_0\sin(k_x x),
\label{eq:psib}
\end{equation}
such that ${\bf u}_{\rm b} = (u_{\rm b},v_{\rm b},0)$ where  
\begin{equation}
    u_{\rm b} = \sin(y) + a u_{\rm m} = - \frac{\partial \psi_{\rm b}}{\partial y}, \quad v_{\rm b} = a v_{\rm m} = \frac{\partial \psi_{\rm b}}{\partial x}. 
\end{equation}
\pg{We see that this flow can be interpreted as a Taylor-Green vortex array, to which two Kolmogorov flows in perpendicular directions have been added. This differs from prior works on the linear stability of vortex arrays \citep[e.g.][]{Suzukial2018,Hattorial2021,Guoal2024,Guoal2025}, which only had the Taylor-Green component of the flow.} 

Figure \ref{fig:meandervisu} shows visualizations of $\bu_{\rm b}$ \pg{for different values of $a$.}  Contours of $\psi_{\rm b}(x,y)$
are shown as colored lines,  overlaid on a colormap of $|{\bf u}_{\rm b}| = \sqrt{u_{\rm b}^2 + v_{\rm b}^2}$. As expected, we see that ${\bf u}_{\rm b}$ exhibits meanders whose amplitude increases with $a$. For $a \ge |v_0^{-1}|/2 = 0.55$, the perturbation amplitude exceeds that of the mean flow $\bar {\bf u}$, and ${\bf u}_{\rm b}$ begins to resemble a vortex array.   

\begin{figure}
\centerline{\includegraphics[width=\textwidth]{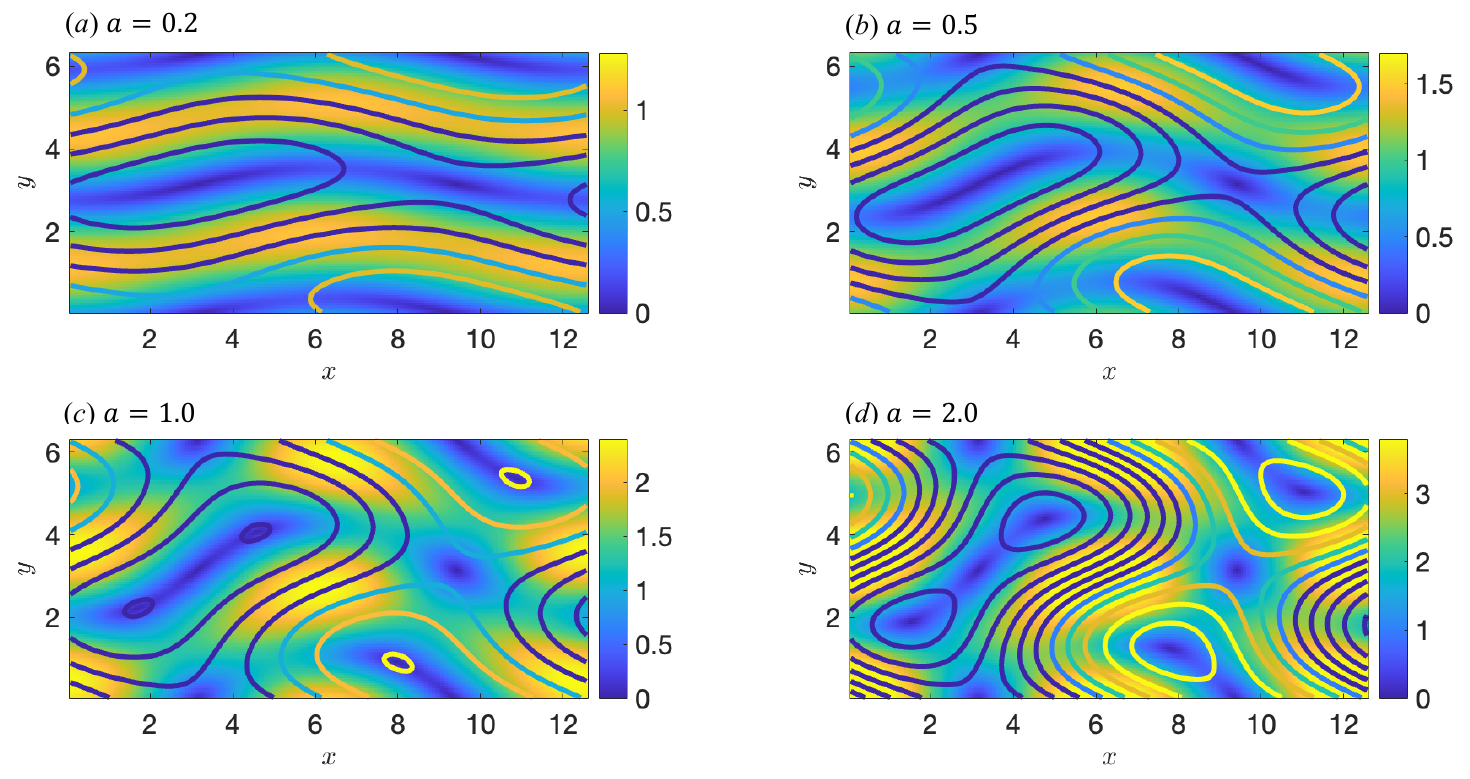}}
\caption{Contour lines of the stream function  $\psi_{\rm b}$  representing the 
meandering flow
$ {\bf u}_{\rm b}$, overlaid on top of color maps of $| {\bf u}_{\rm b}|$ (see equation \ref{eq:ub}), for increasing meander amplitude $a$ from panel ({\it a}) to ({\it d}). Note that the color range is different in the top row and bottom row, and that the colors of the streamlines are arbitrary.}
\label{fig:meandervisu}
\end{figure}

\subsection{Stability analysis of ${\bf u}_{\rm b}$.}
\label{sec:secondarytheory}

We now study the stability of the \pg{`frozen-in'} columnar flow ${\bf u}_{\rm b}$ given in equation (\ref{eq:ub}) to three-dimensional perturbations $\bu'$. We shall refer to these perturbations as `secondary instabilities' of $\bar {\bf u}$. The method used to compute the eigenmodes ${\bf u'}$ and their growth rate is well-established \citep{POTYLITSIN_PELTIER_1998,Deloncleal2011,Hattorial2021,Guoal2024}. The total flow is now given by 
\begin{equation}
    \bu(x,y,z,t) = \bu_{\rm b}(x,y) + \bu'(x,y,z,t).
\end{equation}
The governing equations, linearized around $\bu_{\rm b}$ \pg{and using the frozen-in approximation}, satisfy 
\begin{eqnarray}
\frac{\partial \bu'}{\partial t} + \bu' \cdot \nabla \bu_{\rm b} + \bu_{\rm b} \cdot \nabla \bu' = - \nabla p' + \frac{b'}{Fr^2} \be_z + \frac{1}{Re} \nabla^2 \bu' \label{eq:momsecond}, \\
\frac{\partial b'}{\partial t} + \bu_{\rm b} \cdot \nabla b' + w' = \frac{1}{Pe} \nabla^2 b' ,\\
\nabla \cdot \bu' = 0. \label{eq:divsecond} 
\end{eqnarray} 
Since $\bu_{\rm b}$ depends on $x$ and $y$, the ansatz for the perturbations takes the form
\begin{equation}
q' = \hat q(x,y) e^{i k_z z + \sigma t},
\label{eq:2Dansatz}
\end{equation}
where $q \in \{ u,v,w,b,p\}$. Substituting (\ref{eq:2Dansatz}) into (\ref{eq:momsecond})--(\ref{eq:divsecond}), and expanding the equations into components, we obtain
\begin{eqnarray}
\sigma\hat u + \hat \bu \cdot \nabla_{\rm h} u_{\rm b} + \bu_{\rm b} \cdot \nabla_{\rm h} \hat u  = - \frac{\partial \hat p}{\partial x}  + \frac{1}{Re}\left( \nabla_{\rm h} ^2 \hat u - k_z^2 \hat u\right), \\
\sigma \hat v + \hat \bu \cdot \nabla_{\rm h} v_{\rm b} + \bu_{\rm b} \cdot \nabla_{\rm h} \hat v  = - \frac{\partial \hat p}{\partial y} + \frac{1}{Re} \left( \nabla_{\rm h}^2 \hat v  - k_z^2 \hat v \right), \\
\sigma \hat w + \bu_{\rm b} \cdot \nabla_{\rm h} \hat w  = - i k_z \hat p + \frac{\hat b}{Fr^2} + \frac{1}{Re} \left(\nabla_{\rm h}^2 \hat w  - k_z^2 \hat w \right), \\
\sigma \hat b + \bu_{\rm b} \cdot \nabla_{\rm h} \hat b + \hat w = \frac{1}{Pe} \left(\nabla_{\rm h}^2 \hat b  - k_z^2 \hat b \right), \\
\nabla_{\rm h} \cdot \hat \bu + i k_z \hat w  = 0,
\end{eqnarray}
where $\nabla_{\rm h} = (\partial/\partial_x, \partial/\partial_y,0)$. Having chosen $\bu_{\rm b}$ \pg{and our computational domain} to be
periodic in $x$ with period $L_x = 4\pi$, and in $y$ with period $L_y = 2\pi$, we assume that the perturbations have the same periodicity, and thus seek solutions of the form
\begin{equation}
\hat q(x,y) = \sum_{m=-M}^{M} \sum_{n=-N}^{N} q_{mn} e^{i m k_x x + i n y} , 
\end{equation}
where $k_x = 1/2$ and $k_y = 1$ (so it is omitted to simplify the notation).
When solving the problem numerically, the sums are truncated at finite wavenumbers $M$ and $N$, respectively. 
 The set of algebraic equations for the $5 \times (2M+1)\times (2N+1)$ coefficients $q_{mn}$ are derived and presented in  Appendix B. 
  The equations form a  generalized eigenvalue problem of the form ${\mathbf A} \mathbf X = \sigma \mathbf{B} \mathbf{X}$, where the matrix ${\mathbf A}$ has  complex entries. This problem
 can be solved using standard packages such as the `eig' routine in Matlab. We successfully cross-validated the results with the LAPACK routine ZGGEV, but the latter is not as stable as Matlab's routine when used in double-precision so we used `eig'for all computations. We ensured that the truncation thresholds $M$ and $N$ are sufficiently large to resolve the solution, noting that a lower threshold is required to get an accurate estimate of the growth rate than to get an accurate representation of the entire eigenmode (see Appendix B for detail).

\subsection{Growth rate of the secondary instability}
\label{sec:results}

For each set of input parameters $Re$, $Pe$, $Fr$, and $a$ (the meander amplitude), 
we use the method described in  Appendix B to compute the growth rate $\lambda(k_z)= \Re(\sigma(k_z))$ of the secondary modes of instability. For a given $k_z$, several different modes can exist \citep{Hattorial2021}, but we are only interested in the fastest-growing one.
Figure \ref{fig:results1}({\it a}) presents these growth rates for varying meander amplitude $a$, at fixed $Re = Pe = 10000$, $Fr = 0.1$.
For each data point in the figure, we used $M = 20, N = 10$ and 
 verified that this Fourier resolution is sufficient to obtain a good estimate for $\lambda$ (see Appendix B). 
The case with $a = 0$ recovers the solution for the instability of the basic flow $\bar {\bf u}$ \pg{\citep[see Appendix A and][]{Cocusseetal2025}, where the fastest growing mode has $k_z = 0$}. As  the meander amplitude $a$ increases, the growth rate of the fastest-growing secondary mode of instability also increases. For small $a$ (e.g. $a = 0.2$ and $a = 0.35$), this is primarily due to the overall increase in the background flow velocity. However, for $a \ge 0.5$ we see that the peak of the growth rate curve shifts to a value of $k_z > 0$, indicating the presence of a new instability mechanism. 

\begin{figure}
\includegraphics[width=\textwidth]{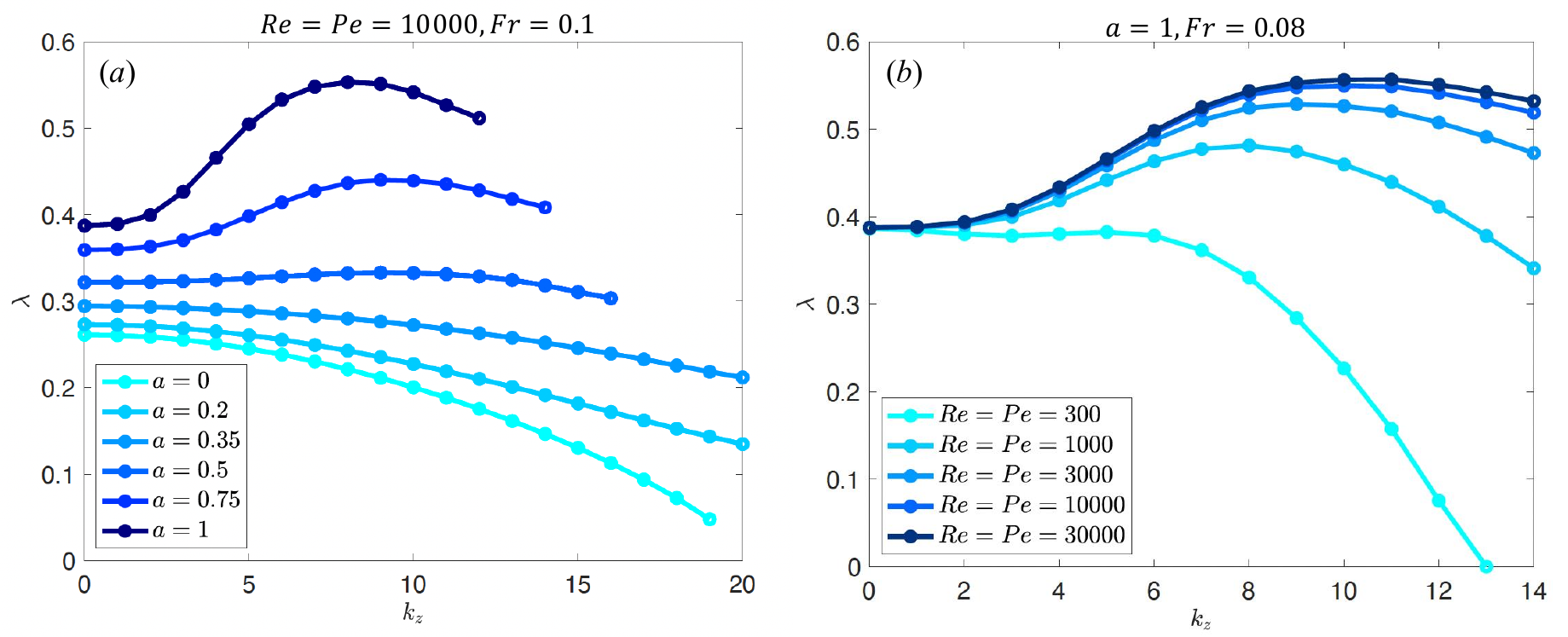}
    \caption{({\it a}) Growth rate of the secondary mode of instability as a function of $k_z$, for $Re = Pe = 10000$, $Fr = 0.1$, and varying $a$ (see legend for detail). In each case $M = 20, N = 10$. ({\it b}) As in ({\it a}), but fixing $a = 1$ and $Fr = 0.08$, and varying $Re$ (with $Pe = Re$). In both panels, we have checked that  $\lambda$ has converged given this Fourier resolution (see Appendix B for detail).} 
\label{fig:results1}
\end{figure}

Figure \ref{fig:results1}({\it b}) shows similar results, but this time fixing $a = 1$, $Fr = 0.08$, and varying $Re$ (with $Pe = Re$). We see that the growth rate of the high $k_z$ modes decreases as  viscosity and thermal diffusivity both increase, which is not surprising. This effect shifts the peak of the $\lambda(k_z)$ curve downwards and towards lower $k_z$, until it disappears completely. In what follows, we always select $Re = Pe = 10000$ \pg{(consistent with the DNS in \S\ref{sec:DNS}). With these values, the effects of viscosity and buoyancy dissipation are negligible for modes with $k_z = O(Fr^{-1})$}.  

Finally, we note by comparing the $a = 1$ curve in figure \ref{fig:results1}({\it a}) to the $Re = 10000$ curve in figure \ref{fig:results1}({\it b}) (two cases that differ only in the  Froude number used) that the peak of the $\lambda(k_z)$ curve shifts to higher $k_z$ for lower $Fr$. We now study this effect more systematically. 

\subsection{Dependence on the stratification}

Figure \ref{fig:data}({\it a}) presents the growth rates of the secondary instability as a function of $k_z$  for varying Froude number $Fr$, while fixing $Re = Pe = 10000$, and $a = 1$. We define $\lambda_{\rm max} = \max_{k_z} \lambda(k_z)$, and $k_{\rm max}$ as the wavenumber at which that maximum is achieved.
We see that $k_{\rm max}$ increases substantially with  stratification (i.e. as $Fr$ decreases), but that $\lambda_{\rm max}$ is only mildly affected by it. 

\begin{figure}
\includegraphics[width=\textwidth]{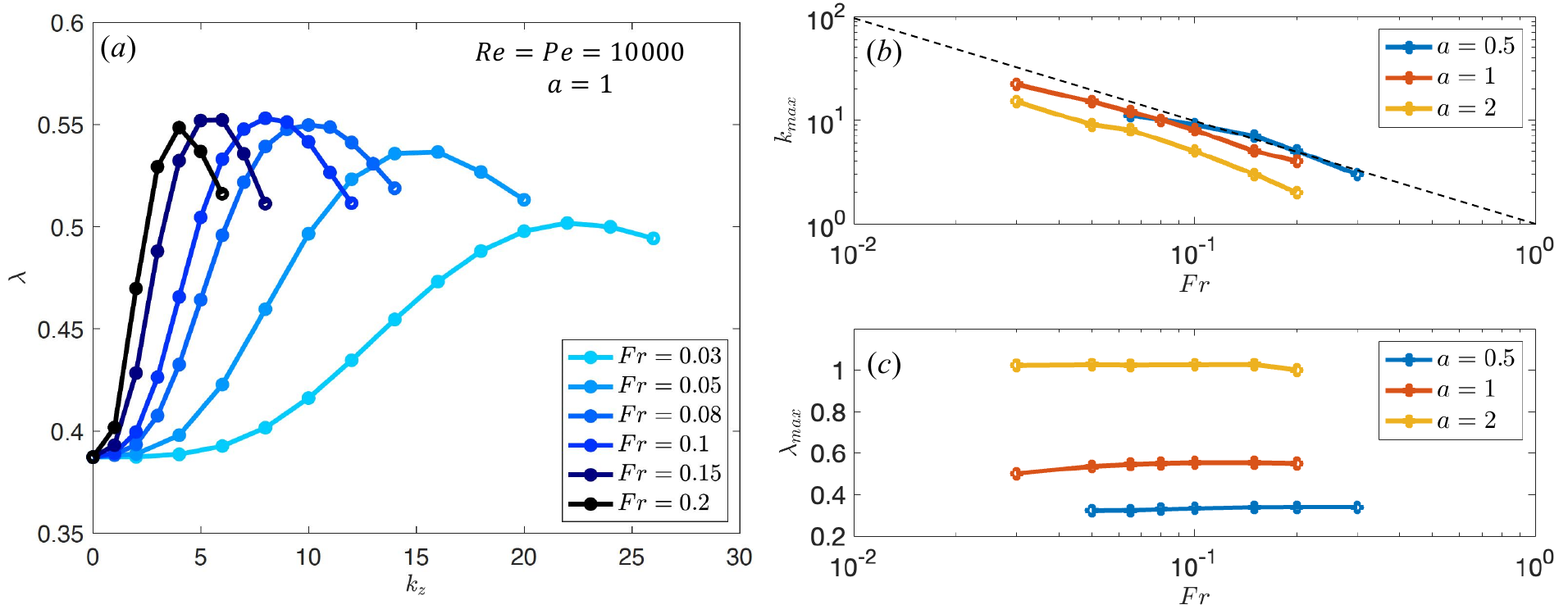}
    \caption{({\it a}) Growth rate of the secondary instability  $\lambda(k_z)$ for $Re = Pe = 10000$, $a=1$, $M=20$, and varying $Fr$ (see legend). 
     ({\it b}) Wavenumber $k_{\rm max}$ at which $\lambda(k_z)$ peaks, as a function of $Fr$, for $Re = Pe = 10000$ and three different values of $a$ (see legend). The dashed line is the line $k_{\rm max} = Fr^{-1}$. ({\it c}) Growth rate $\lambda_{\rm max} = \max_{k_z} \lambda(k_z)$ as a function of $Fr$, for the same parameters as panel ({\it b}). } 
\label{fig:data}
\end{figure}

Figure \ref{fig:data}({\it b}) shows $k_{\rm max}$ and figure \ref{fig:data}({\it c}) shows the corresponding growth rate $\lambda_{\rm max}$ of the fastest-growing mode of secondary instability for $Re = Pe = 10000$, and three different values of $a$, as functions of $Fr$. This confirms that $\lambda_{ max}$ is more-or-less independent of $Fr$ and that $k_{\rm max}$ scales as $Fr^{-1}$ (dashed line), with some dependence on the meander amplitude $a$ for larger values of $a$. These results are qualitatively consistent with those of previous work on the stability of regular stratified vortex arrays \citep{Deloncleal2011, Suzukial2018,Hattorial2021}, and especially those of \citet{Guoal2024} who report very similar scaling laws for the stability of a Taylor-Green  array of vortices. This suggests that the instability mechanism for these regular arrays likely has the same nature as for the flow ${\bf u}_{\rm b}$ studied here. We now briefly review recent results on this topic. 

\citet{Deloncleal2011} studied the stability of columnar vortex arrays resembling von Karman vortex streets to three-dimensional perturbations, and identified the zigzag instability \citep{BillantChomaz2000_experiment} in the limit of strong stratification. \citet{Suzukial2018}  studied the stability of Taylor-Green vortex arrays, as well as Stuart vortices, and argued that the dominant modes of instability at low Froude number are stratified hyperbolic instabilities, that crucially rely on the existence of a hyperbolic point in the flow. \citet{Hattorial2021} investigated the same flows and further classified different types of hyperbolic instabilities (pure hyperbolic, strato-hyperbolic, and mixed hyperbolic) based on the properties of their eigenmodes.  Finally, \citet{Guoal2024} studied the linear stability and nonlinear evolution of a Taylor-Green vortex array. They recovered the results of \citet{Hattorial2021}, and  argued that their `mixed hyperbolic' mode of instability shares key features with the zigzag instability, and is generated through a similar excitation mechanism. \pg{The classification of hyperbolic instabilities proposed by \citet{Hattorial2021} seems specific to the Taylor-Green vortex array, however, and it is not clear how to extend it to other flows.} 

In any case, the vortex array that emerges in our DNS as $a$ increases has a different structure from those considered in previous studies, so prior results cannot be directly applied here. It is therefore necessary to examine the structure of the unstable modes in detail, in order to identify the nature of the high $k_z$ secondary instability of the flow ${\bf u}_{\rm b}$.  

   \subsection{Structure of the fastest-growing eigenmode}

The structure of the secondary modes of instability for $Re = Pe = 10000$, $Fr = 0.1$ and varying $a$ is illustrated in figure \ref{fig:Modecompare}. The left column shows the background vorticity $\omega_{\rm b} = (\nabla \times \bu_{\rm b})\cdot \be_z$ as a color map, overlaid with contours of $\psi_{\rm b}$, while the middle and right columns shows $\omega'_{z}$ and $w'$, respectively, for the selected eigenmode. For $a =0$ to $a = 0.75$, we show the mode with $k_z = 9$ -- this is the value of $k_z$ of the fastest-growing mode for $a=  0.5$ and $a = 0.75$. \pg{By keeping $k_z$ fixed, we therefore focus on the impact of varying $a$ only.} For
$a =1$ and $a =2$, we show the mode with $k_z = k_{\rm max}$ (namely,  $k_z = 8$ for $a = 1$, and $k_z = 5$ for $a = 2$, respectively). To ensure that the  modes are in the same phase in each panel, we have  set the Fourier coefficient $u_{11} = 1$. The modes are shown at $z = \pi/(4 k_z)$, which is a height at which both $\omega'_z$ and $w'$ have non-zero amplitudes for this selected phase. 
In the middle and right columns, we also show contours of $\psi_{\rm b} = 0$ in white (including arrows to show the flow direction) and of $\omega_{\rm b} = 0$ in black. 

\begin{figure}
\includegraphics[width=\textwidth]{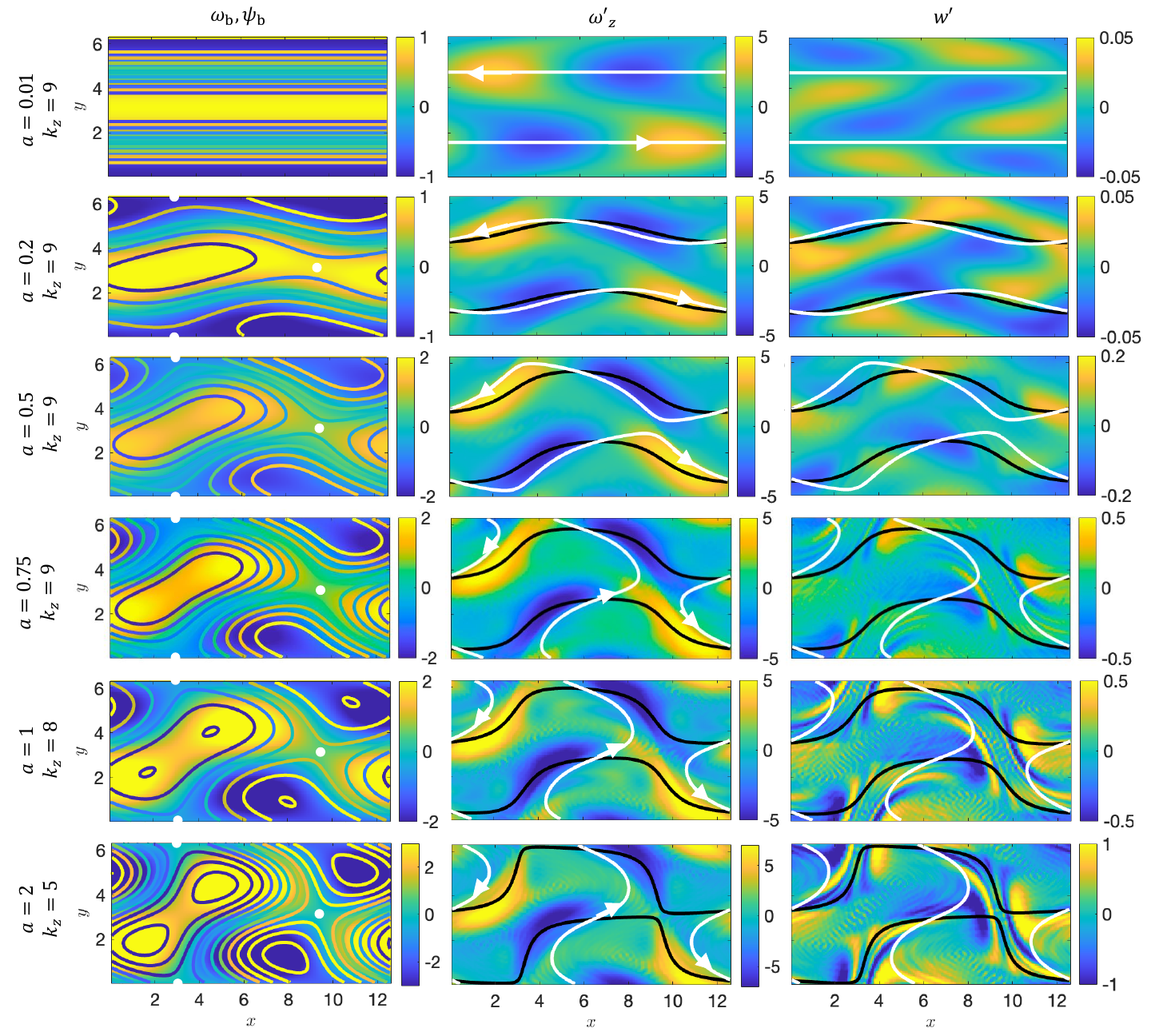}
    \caption{Left column: The background vorticity field $\omega_{\rm b}$ is shown as a color map, and the background streamfunction $\psi_{\rm b}$ is overlaid as a contour map. The values of the meander amplitude $a$ used for each map is shown to the left of the panel. The corresponding structure of the eigenmode for $\omega'_z$ and $w'$, computed for $Re = Pe = 10000, Fr = 0.1$ are shown in the center and right columns, respectively. The value of $k_z$ used is shown on the left. The Fourier resolution is $M = 10$ for $a = 0.01$, $M = 20$ for $a = 0.2$, $M = 40$ for $a = 0.5$, and $M = 50$ for $a \ge 0.75$. In each case $N = M/2$. The white contours in the middle and right columns show $\psi_{\rm b} = 0$, and the black contours show $\omega_{\rm b} = 0$. \pg{The white dot in the left column indicates the position of the primary hyperbolic points in the flow, at $(3\pi,\pi)$ and ($\pi,0$).}}
    \label{fig:Modecompare}
\end{figure}

The mode structure in the absence of meander (for $a = 0.01$, top row) for $k_z = 9$ is shown in the top row. It primarily consists of large-scale vortices that are somewhat elongated in the streamwise direction, and change sign across the base flow inflection points (at $y  =0$ and $y = \pi$). \pg{The vertical vorticity $\omega_z'$} resembles that of the $k_z = 0$ mode shown in Appendix A (figure \ref{fig:vorticitymap}{\it b}), \pg{as expected from the analysis of \citet{Cocusseetal2025}.} The vertical flow $w'$ has a simple chevron structure that changes sign at the values of $y$ where $\psi_{\rm b} = \omega_{\rm b} = 0$.

When the meander amplitude $a$ is small ($a = 0.2$), we recall from figure \ref{fig:results1}({\it a}) that the growth rate curve $\lambda(k_z)$ is very similar to that of the primary instability, with a maximum at $k_z =0$.  Inspection of the mode structure  (see figure \ref{fig:Modecompare}, second row) shows that it looks very similar to the $a = 0.01$ eigenmode, with an added undulation presumably due to the advection by the background flow ${\bf u}_{\rm b}$. The same is true of the $a = 0.5$ case (third row in figure \ref{fig:Modecompare}), but the undulation is more pronounced.

When $a$ increases to $0.75$,  $\lambda(k_z)$ now peaks at $k_z \simeq 9$ instead of $k_z = 0$ in figure \ref{fig:results1}({\it a}). \pg{This was interpreted earlier as implying the existence of a new mode of instability. The fourth row of figure \ref{fig:Modecompare} confirms this hypothesis: we see the emergence of} fine 'ripples' in $w'$ (and $b'$, not shown), which are centered around the flow's primary hyperbolic points at $x = 3\pi, y = \pi$, \pg{and $x = \pi, y =0$} (\pg{shown in the white dots on figure \ref{fig:Modecompare}, left column}). The ripples are more-or-less perpendicular to the convergent manifold of that point, and parallel  with the diverging manifold. The cases with $a = 1$ and $a = 2$ are very similar, with prominent ripples in $w'$ and $b'$ around the primary hyperbolic points \pg{at ($3\pi,\pi$) and ($\pi,0$)}.  Additional faint structures can be seen around a second hyperbolic point at $x = \pi, y =  \pi$, and around the centers of each vortex, but as they are not well resolved even with $M = 50$, we do not discuss them further. 

\subsection{Clues to the hyperbolic instability}
\label{sec:hyperbolic}

The obvious role of the background flow's hyperbolic points in shaping the secondary mode structure at large values of  $a$ strongly suggests the presence of a hyperbolic instability. In this section, we provide further evidence supporting this interpretation. 

The unstratified inviscid hyperbolic instability \pg{around an isolated hyperbolic point} was investigated \pg{in detail} by \citet{KerswellCaulfield2000}. \pg{Using a specific ansatz consisting of a plane wave whose amplitude and wave vector are allowed to evolve with time only \citep{lifschitz1991local,FriedlanderVishik1991},} they showed 
 that such perturbations can grow exponentially at a rate $\Delta_{\rm h}$, where 
\begin{equation}
   \Delta_{\rm h} = \Delta({\bf x}_{\rm h}) \mbox{ with }   \Delta^2({\bf x}) =  \frac{\partial u}{\partial y}\frac{\partial v}{\partial x} + \left(\frac{\partial v}{\partial y}\right)^2 , 
    \label{eq:Delta2}
\end{equation}
and where ${\bf x}_{\rm h}$ is the \pg{position of the} flow's hyperbolic point. The quantity $\Delta^2$ is positive around a hyperbolic point, and negative around an elliptic point. \pg{While the type of perturbation and background flow considered by \citet{KerswellCaulfield2000} is highly restrictive, \citet{RajkotiaZaheeretal2025} have recently shown that the same results can be recovered for a more general multiscale ansatz using a WKB analysis: appropriately oriented small-scale, rapidly varying perturbations are simply advected by the 2D flow field towards and then away from the hyperbolic point, undergoing amplification in the process. This `local' linear stability analysis naturally generalizes to any flow with a hyperbolic point, isolated or not, providing an estimate of the instantaneous growth rate of the perturbation following its trajectory in the flow.}  

For the background flow ${\bf u}_{\rm b}$ considered here (see equation \ref{eq:ub}), it is easy to show that at the primary hyperbolic point ${\bf x}_{\rm h} = (3\pi,\pi)$, 
\begin{equation}
\Delta_{\rm h} = \sqrt{ a^2 -\frac{av_0}{2}},
\label{eq:Delta}
\end{equation}
where we recall that $v_0 = -0.913$. Figure \ref{fig:hyperbolic} shows $\Delta_{\rm h}$ as a function of $a$ (blue solid line). It also shows the line $\lambda = 0.26$ (horizontal green dashed line), which is the growth rate of the primary instability for $k_x = 0.5$, and $k_z = 0$  at $Re = Pe = 10000$. The growth rates 
of the fastest-growing secondary modes of instability
$\lambda_{\rm max} = \max_{k_z}(\lambda)$ for varying $a$, at fixed values of $Re = Pe = 10000$ and several values of $Fr$ are shown as a black line and colored symbols (see legend for detail).  

 We see that $\lambda_{\rm max}$ is equal to the growth rate of the fastest-growing primary mode at low meander amplitudes (small $a$), and is approximately equal to $0.46 \Delta_{\rm h}$ for large $a$ (dashed blue line), regardless of stratification. The 0.46 prefactor was fitted to the data. 
The fact that $\lambda_{\rm max} < \Delta_{\rm h}$ is not surprising, as the analysis of \citet{KerswellCaulfield2000} assumes a single isolated hyperbolic point, which is not the case here. The background flow ${\bf u}_{\rm b}$ also has elliptic points where $\Delta^2$ is negative, around which perturbations are oscillatory rather than amplified. A localized perturbation advected by the streamlines  therefore spends part of its time being amplified  in the vicinity of the hyperbolic point, and part of its time oscillating around the elliptic point \citep{Suzukial2018}, which explains why  $\lambda_{\rm max} < \Delta_{\rm h}$. 
Nevertheless, it is quite remarkable that the simple relation $\lambda_{\rm max} = 0.46 \Delta_{\rm h}$ fits so well for sufficiently large $a$, regardless of $Fr$, despite the fact that our hyperbolic point is stratified while the one studied by \citet{KerswellCaulfield2000} is not. 

Finally, we note that figure \ref{fig:hyperbolic} also clarifies why the ripples in $w'$
 only appear for $a > 0.5$ in figure \ref{fig:Modecompare}. That is precisely the point at which the hyperbolic instability growth rate $\lambda = 0.46 \Delta_{\rm h}$ starts exceeding the growth rate of  the primary instability $\lambda = 0.26$.

\begin{figure}
\begin{center} \includegraphics[width=0.65\textwidth]{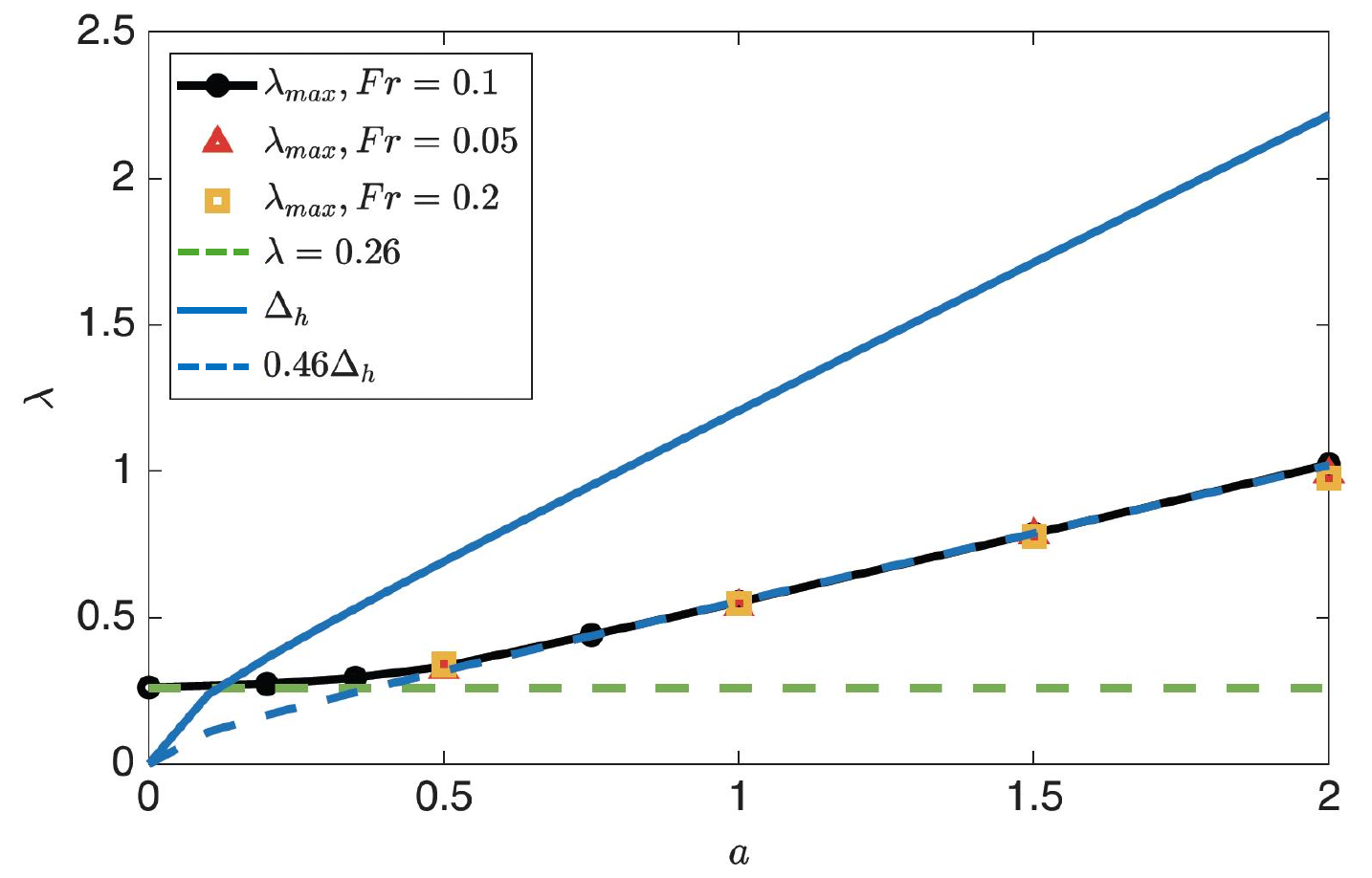}
\end{center}
    \caption{Symbols show the growth rate $\lambda_{\rm max}$ of the fastest-growing mode of secondary instability  as a function of meander amplitude $a$, for $Re = Pe = 10000$ and different values of $Fr$ (see legend). Also shown is the growth rate of the fastest-growing mode of the primary instability ($\lambda = 0.26$, green dashed line), the growth rate of the hyperbolic instability \pg{for an isolated hyperbolic point} $\Delta_{\rm h}$ (blue line, see equation \ref{eq:Delta}) as well as $0.46\Delta_{\rm h}$ (blue dashed line), which fits the actual growth rates well for larger $a$.}
    \label{fig:hyperbolic}
\end{figure}

\subsection{Comparison with Direct Numerical Simulations}
\label{subsec:DNScomp}

\pg{While the analysis presented so far generically looked at the linear stability of columnar flows of the form ${\bf u}_{\rm b}(x,y)$ given in (\ref{eq:ub}) and (\ref{eq:psib}), this only approximates the 2D flow in DNS2 at early times, prior to the saturation of the primary instability (around $t=10$). However, the secondary modes only begin to grow later (from $t = 40$ to $t = 100$). We now use a new} background flow $\bu_{{\rm b,DNS}}$ that more closely mimics the one observed in DNS2 \pg{during the growth of the secondary instability. To construct it, we select a particular time $t = 84$. At this time, the flows in DNS2a $(Fr = 0.15)$ and DNS2b $(Fr = 0.1)$ are still almost identical, and essentially vertically invariant.  A snapshot of  $\omega_z(x,y)$ at $z = 0$ and $t = 84$ is shown in figure \ref{fig:Datavstheory}({\it a}).}  We seek a plausible  approximation of \pg{the corresponding 2D flow of} the form
\begin{equation}
\bu_{{\rm b,DNS}} = \left[ U_0 \sin(y) + A k_x^{-1} \cos(y) \cos(k_x x)   \right] \be_x + \left[ V_0 \cos(k_x x) + A \sin(y) \sin(k_x x) \right] \be_y,  
\label{eq:ubDNS}
\end{equation}
which \pg{is again equal to a Taylor-Green vortex array plus two sinusoidal flows in perpendicular directions. This} enables us to leverage the tools and intuition developed in the previous sections. A rigorous way of identifying the coefficients $V_0$ and $U_0$ in (\ref{eq:ubDNS}) is to project the vertical vorticity $\omega_z$ at $t = 84$ onto the corresponding Fourier modes, to obtain
\begin{eqnarray}
U_0 = - \frac{2}{L_xL_y} \int_0^{L_x} \int_0^{L_y} \omega_z \cos(y) dxdy \simeq 0.045, \\
V_0 = - \frac{2}{k_x L_xL_y} \int_0^{L_x} \int_0^{L_y} \omega_z \sin(k_x  x) dxdy \simeq -0.93.
\label{eq:Aeq}
\end{eqnarray}
\pg{We then fit the remaining coefficient 
\begin{equation}
    A = 0.4 
\end{equation} 
to match, approximately, the amplitude of $\omega_z$ within the vortices in figure \ref{fig:Datavstheory}({\it a}). }

\begin{figure}
    \includegraphics[width=\textwidth]{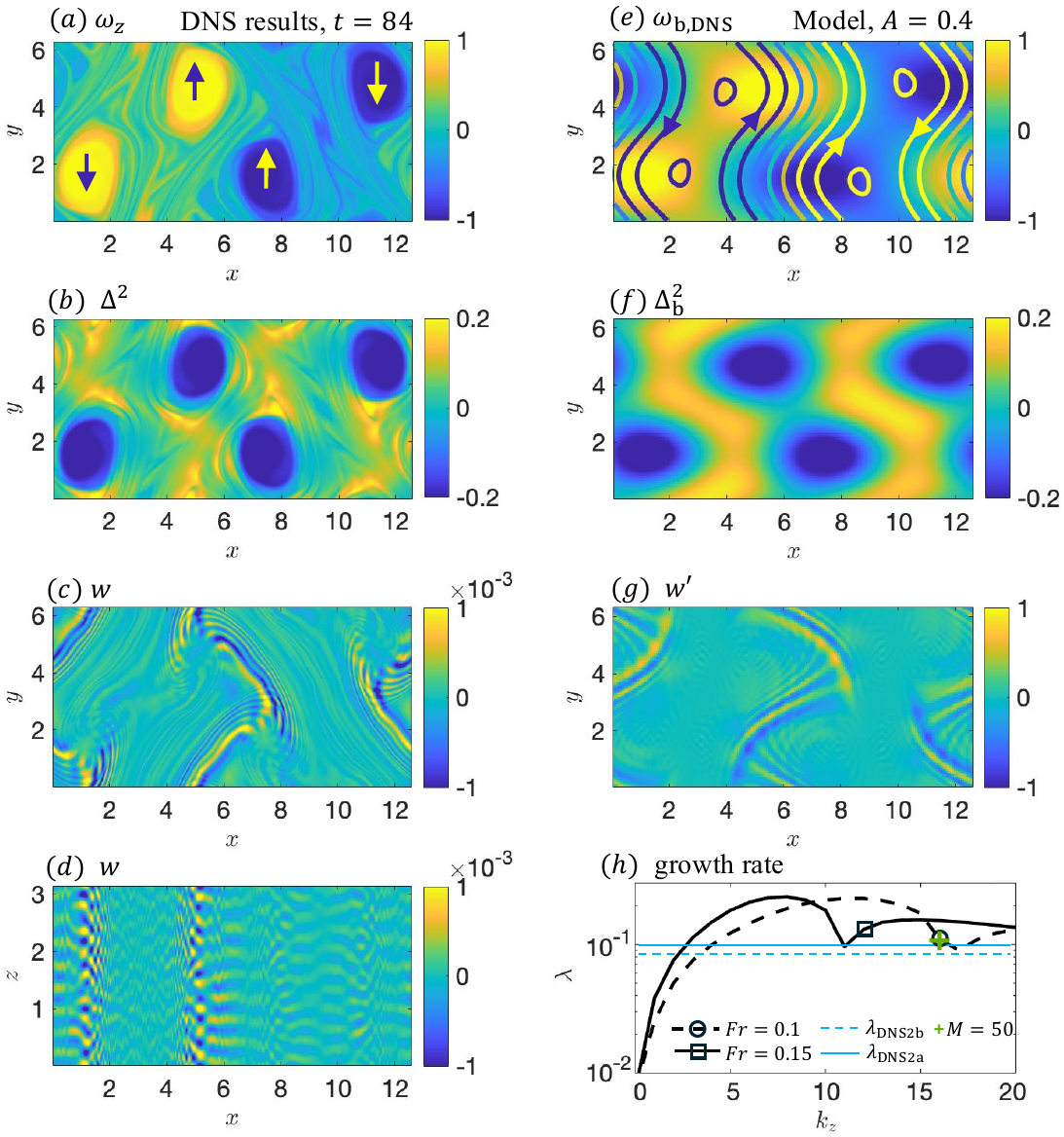}
    \caption{Comparison between the results of DNS\pg{2b} ($Re = Pe = 10000$, $Fr = 0.1$), in the left column, with linear theory for the secondary instability of the steady flow $\bu_{\rm b,DNS}$  given in equation (\ref{eq:ubDNS}) with $A = 0.4$, $V_0 = -0.93$, $U_0 = 0.045$ in the right column. ({\it a,c}) DNS snapshots of  $\omega_z$ and $w$, respectively, at $t = 84$ in the $z= 0$ plane. ({\it b}) $\Delta^2$ (see equation \ref{eq:Delta2}) for the same snapshot. ({\it d}) Snapshot of $w$ at the same time on the $y =0$ plane. \pg{The vertical axis has been stretched to better visualize the flow.}  
    ({\it e}) Model vorticity field $\omega_{{\rm b,DNS}}$. ({\it f}) $\Delta_{\rm b}^2$ for the model flow. ({\it g}) Structure of the $w'$ component of the linear eigenmode of secondary instability for \pg{$Fr = 0.1$, $Re = Pe = 10000$, and} $k_z = 16$, which is the dominant secondary mode observed in DNS2b. The amplitude is arbitrary.  ({\it h}) Growth rate $\lambda(k_z)$ for the secondary mode of instability for the flow ${\bf u}_{{\rm b,DNS}}$ with \pg{$Re = Pe = 10000$,  for $Fr = 0.1$ (dashed line) and $Fr = 0.15$} (solid line). \pg{The growth rates of the modes observed to grow in the DNS ($k_z = 16$ with $\lambda_{{\rm DNS2b}} = 0.084$ for $Fr = 0.1$, and $k_z = 12$ with $\lambda_{{\rm DNS2a}}=0.1$ for $Fr = 0.15$)} are shown in the blue lines, \pg{and the linear stability theory predictions for the same modes are shown in the symbols}. The Fourier resolution used is $M = 30$ for most points, and $M=50$ was used to confirm the results for the $k_z = 16$ point (green $+$ sign), \pg{and to create panel ({\it g})}. The arrows in ({\it a}) and ({\it e}) show the direction of the flow in the $y$ direction.} 
\label{fig:Datavstheory}
\end{figure}

 The corresponding vertical vorticity of the model flow $\omega_{{\rm b,DNS}}$ is shown in figure \ref{fig:Datavstheory}({\it e}) for comparison. We also compare $\Delta^2_{\rm b}(x,y)= (\partial u_{\rm b}/\partial y)( \partial v_{\rm b}/\partial x) + (\partial v_{\rm b} /\partial y)^2$ of the model flow (\ref{eq:ubDNS}) in figure \ref{fig:Datavstheory}({\it f}) to $\Delta^2(x,y)$ of DNS2 at $t = 84$ in figure \ref{fig:Datavstheory}({\it b}). With $A = 0.4$, both have roughly the same negative value within the vortices. The same applies close to the hyperbolic points, with the maximum values of $\Delta_{\rm b}^2$ and $\Delta^2$ both being positive and approximately equal to $0.2$. While the large-scale structure of $\Delta^2$ is well captured by the model, we note that DNS2 also shows finer-scale features, which is expected given that the streamlines -- and thus the hyperbolic points -- in the DNS are not steady. 

We then perform a linear stability analysis of the 2D flow ${\bf u}_{{\rm b,DNS}}(x,y)$ to 3D perturbations using the method described in Appendix B. \pg{We first look at DNS2b, for which $Re = Pe = 10000$ and $Fr = 0.1$.}  
 Figure \ref{fig:Datavstheory}({\it h}) shows the predicted growth rate of the fastest-growing secondary modes of instability for varying $k_z$ \pg{(dashed line), at these parameters. Interestingly, 
 we find that the fastest-growing mode has $k_z \simeq 12$, which is significantly smaller than the vertical wavenumber of the mode actually observed to grow in DNS2b, which has $k_z = 16$, see \S\ref{subsec:DNS2} and figure \ref{fig:Datavstheory}({\it d}). Intriguingly, we find that the  $k_z = 16$ mode is actually close to a local minimum of the $\lambda(k_z)$ curve, see the green + symbol in figure \ref{fig:Datavstheory}({\it h}), which raises the (unanswered) question of why it is actually preferred in DNS2b.}
 Nevertheless,
  we also find that the growth rate of the $k_z = 16$ perturbation is approximately $0.1$, which is quite close to the empirically-determined one ($\lambda_{{\rm DNS2b}} \simeq 0.084$, see dashed blue line in figure \ref{fig:Datavstheory}({\it h})), and also close to $0.46 \Delta_{\rm h}$ (where $\Delta_{\rm h} \simeq 0.2$), as in \S\ref{sec:hyperbolic}, thus demonstrating this scaling is robust. Finally, the  vertical velocity component $w'$ of the $k_z = 16$  mode is shown in figure 
\ref{fig:Datavstheory}({\it e}). We see that even though $\omega_z$ is time-dependent, and only crudely modeled by $\omega_{\rm b,DNS}$, many of the salient features of $w$ \pg{in DNS2b} are correctly captured by $w'$ (e.g. compare figure \ref{fig:Datavstheory}({\it g}) with \ref{fig:Datavstheory}({\it c})).   

\pg{We also computed the growth rate of 3D perturbations to  ${\bf u}_{{\rm b,DNS}}(x,y)$ for $Fr = 0.15$ (as in DNS2a). }
    The results are shown in the \pg{solid line in figure \ref{fig:Datavstheory}({\it h}). Again, we find that the mode that is actually observed to grow in DNS2a ($k_z = 12$) is not the fastest-growing one, and instead, lies close to a local minimum of the $\lambda(k_z)$ curve. The predicted growth rate of the $k_z = 12$ mode, however, is close to the measured value of $\lambda_{\rm DNS2a} = 0.1$.

    The systematic discrepancy between the wavenumber of the fastest-growing mode and the mode actually observed to grow in the DNS was unexpected, and} could be due to the time-dependence of the background flow (which is ignored when using the frozen-in approximation). \pg{This effect results in the growth of perturbations with somewhat smaller vertical scales than anticipated. Other than}  the problem of mode selection, however, the good qualitative and quantitative match between the predictions from the linear stability analysis and the DNS data is quite remarkable, and thus confirms \pg{that the secondary instability in DNS2 is actually a high $k_z$  hyperbolic instability, with a growth rate close to $\lambda = 0.46 \Delta_{\rm h}$. Classifying the subtype of hyperbolic instability this is (assuming that it can  be classified in the same way as \citet{Hattorial2021} in time-dependent flows) is beyond the scope of this paper.}

\section{\pg{Diapycnal mixing and mixing efficiency}}
\label{sec:eta}

\pg{We finally turn to the second question raised in \S\ref{sec:intro}, regarding the impact of pathway selection on diapycnal mixing and mixing efficiency. In this section, we focus on the $Fr = 0.15$ case for simplicity, although the same results apply to the $Fr = 0.1$ case.  In figures \ref{fig:DNS1}({\it d}) and \ref{fig:DNS2}({\it d}) we saw that the maximum mixing efficiency achieved is $\eta_{\rm max} \simeq 0.25$ for DNS1a and $\eta_{\rm max} \simeq 0.165$ for DNS2a, even though these simulations both have exactly the same input parameters, and more or less the same peak value of $\chi$, see figure \ref{fig:DNS1}({\it c}) and \ref{fig:DNS2}({\it c}). This difference is quite substantial, and needs to be explained. 

One may first wonder whether the effective buoyancy Reynolds number and/or effective Froude number of the flow are very different between the two cases at the time of peak mixing. Indeed, the large-scale horizontal flow evolves and loses energy with time, thereby reducing the effective buoyancy Reynolds number $Re_{b,{\rm eff}}$ and effective Froude number $Fr_{\rm eff}$, defined here as
\begin{equation}
Fr_{\rm eff}(t) = u_{{\rm h,rms}}(t) Fr \mbox{  and } Re_{b,{\rm eff}}(t) = u^3_{\rm h,rms}(t) Re_{b},
\label{eq:freff}
\end{equation}
where $u_{\rm h,rms}(t) = \sqrt{u^2_{\rm rms}(t) + v^2_{\rm rms}(t)}$ is the rms horizontal velocity of the flow. (This assumes that the characteristic horizontal length scale of the large-scale flow does not change significantly with time, which is a good approximation for these DNS.)

Figure \ref{fig:lengthscales}({\it a,b}) shows the evolution of $\langle | \nabla \bu|^2 \rangle (t)$ and $\langle | \nabla b|^2 \rangle (t)$ for DNS1a (purple) and DNS2a (green), as functions of $Re_{b,{\rm eff}}(t)$ (a plot against $Fr_{\rm eff}(t)$ would look qualitatively similar). To avoid crowding the figure, we do not show the early  evolution, focusing instead on times $t > t_{1/4}$, where $t_{1/4}$ marks the time where the area fraction of the $y=0$ plane occupied by regions with $Ri<1/4$ begins to exceed 0.001. This approximately corresponds to the onset of the KH instabilities in each DNS, and is shown as a filled symbol in figures \ref{fig:lengthscales}({\it a,b}). Open symbols in figure \ref{fig:lengthscales}({\it a,b}) mark the times of highest KH activity in DNS1a and DNS2a, defined as times where the area fraction of the $y=0$ plane occupied by regions with $Ri<0$ exceeds 0.001. 

We first see that $Re_{b,{\rm eff}}$ is approximately the same in the two simulations at the time of highest KH activity (the same would be true of $Fr_{\rm eff})$, so this does not explain the difference between the observed peak mixing efficiencies. 
We also see that during that time, $\langle | \nabla b|^2 \rangle$ is essentially the same in DNS1a and DNS2a (confirming what we saw in figures \ref{fig:DNS1}({\it c})) and \ref{fig:DNS2}({\it c}), but $\langle | \nabla \bu|^2 \rangle$ by contrast is significantly larger in DNS2a than in DNS1a. Given the dependence of the mixing efficiency $\eta$ on both quantities, see equations \eqref{eq:etadef} and \eqref{eq:epschidef}, this immediately explains why $\eta_{\rm max}$ is larger in DNS1a than in DNS2a: DNS1a has a substantially lower $\epsilon$, but the same $\chi$, at the same $Re_{b,{\rm eff}}$ and/or the same $Fr_{\rm eff}$.

\begin{figure}
\includegraphics[width=\textwidth]{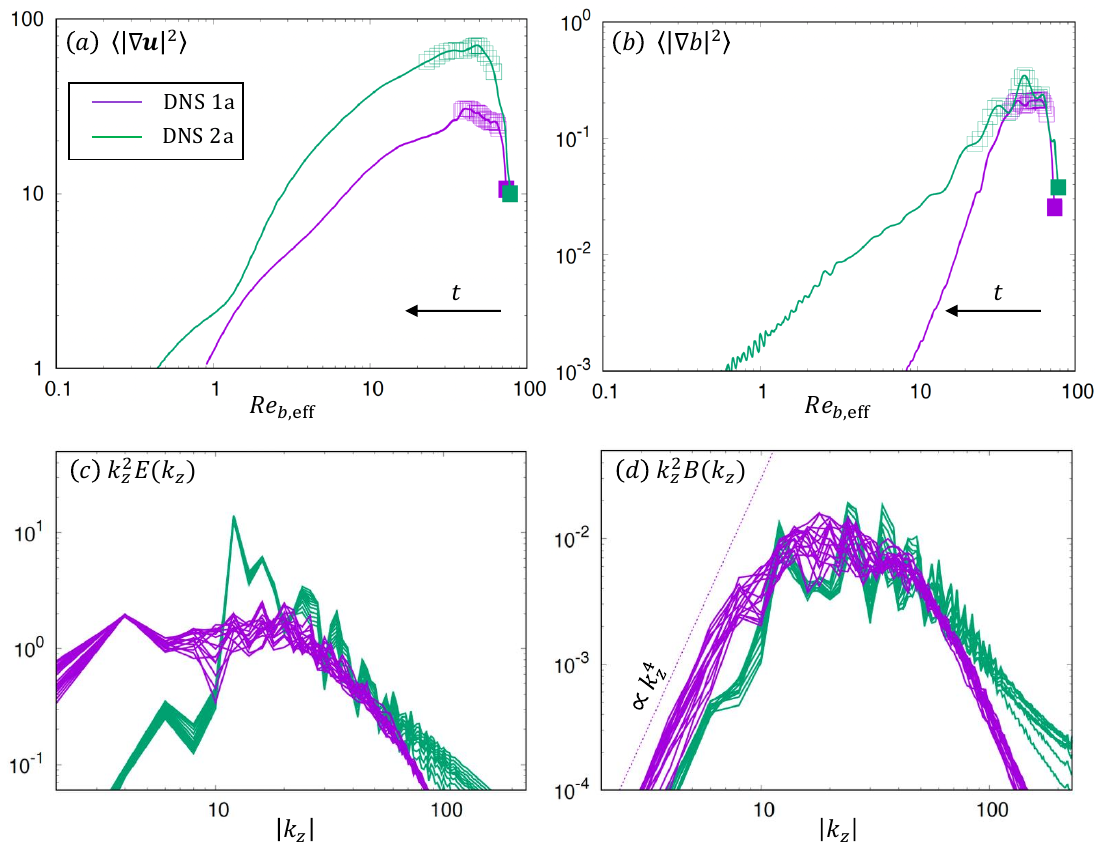}
    \caption{\pg{Comparison of various quantities associated with energy dissipation in DNS1a (purple lines and symbols) and DNS2a (green lines and symbols), where both have $Re = Pe = 10000, Fr = 0.15$). ({\it a}) $\langle |\nabla \bu|^2 \rangle(t)$ against effective buoyancy Reynolds number $Re_{b,{\rm eff}}(t)$ defined in (\ref{eq:freff}), starting from $t = t_{1/4}$, the time when the area fraction of the $y=0$ plane occupied by regions with $Ri<1/4$ first begins to exceeds 0.001. ({\it b}) As in panel ({\it a}), but for $\langle |\nabla b|^2 \rangle(t)$. ({\it c}) Vertical energy dissipation spectra, where $E(k_z)$ is defined in \eqref{eq:vdissspec}, during the time of high KH activity, defined as time where the area fraction of the $y= 0$ plane occupied by $Ri<0$ regions is greater than 0.001. ({\it f}) As in ({\it d}) but for the vertical buoyancy dissipation spectra, where $B(k_z)$ is defined in \eqref{eq:bdissspec}. In panels ({\it a-d}) filled symbols mark $t_{1/4}$ and open symbols mark the times of high KH activity.}}
    \label{fig:lengthscales}
\end{figure}

As we now demonstrate, the significant difference in the viscous dissipation rates observed in the two different pathways to saturation can be attributed to the combination of two factors: (1) the strong anisotropy of the flow, and relatively low volume fraction of regions occupied by KH billows in this strongly stratified limit and (2) the fundamental difference in the range of vertical length scales (or equivalently, wavenumbers $k_z \neq 0$) excited by the primary instability in pathway P1 and the secondary hyperbolic instability in pathway P2. 

The flow anisotropy is clearly visible in figures \ref{fig:DNS1}({\it a})
 and \ref{fig:DNS2}({\it a}): most of the total kinetic energy $0.5 (u_{\rm rms}^2 + v_{\rm rms}^2 + w_{\rm rms}^2)$ is contained in the horizontal flow components at all times. Similarly, the low volume fraction occupied by the billows is shown in figures \ref{fig:DNS1}({\it b})
 and \ref{fig:DNS2}({\it b}).  Consequently, most of the kinetic energy dissipation is due to the viscous dissipation of mostly laminar horizontal flow ($u,v$) in the shallow vertical shear layers that emerge in P1 and P2. 

Figure\ref{fig:lengthscales}({\it c}) shows the vertical dissipation spectrum $k_z^2 E(k_z)$  of  kinetic energy during the time of high KH activity (each open symbol in the other panels corresponding to one spectrum). For comparison, the vertical dissipation spectrum of buoyancy variance is shown in figure\ref{fig:lengthscales}({\it d}). The quantities $E(k_z)$  and $B(k_z)$ are defined as
\begin{eqnarray}
E(k_z) = \sum_{k_x,k_y} |\hat u(k_x,k_y,k_z)|^2 +  |\hat v(k_x,k_y,k_z)|^2 + |\hat w(k_x,k_y,k_z)|^2 \label{eq:vdissspec}, \\
B(k_z) = \sum_{k_x,k_y} |\hat b(k_x,k_y,k_z)|^2,
\label{eq:bdissspec}
\end{eqnarray}
respectively, where hatted quantities in this expression denote their corresponding Fourier coefficients. Note that by plotting $k_z^2 E(k_z)$ and $k_z^2 B(k_z)$, we focus on the dissipation due to vertical gradients of velocity and buoyancy only. This underestimates the total dissipation slightly, especially for DNS2 where the perturbations have small horizontal scales, but does not affect the  conclusions presented below. We see that
the kinetic energy dissipation spectrum is 
essentially flat from low $|k_z|$ up to $|k_z| \simeq 20$ in DNS1a, while it has a strong peak around $k_z = 12$ (which is the wavenumber of the hyperbolic mode identified in \S\ref{subsec:DNS2} and \S\ref{subsec:DNScomp}) in DNS2a. This finding is consistent with the fact that a broad range of vertical scales are excited by the $k_z\neq 0$ modes of the primary instability in pathway P1 (DNS1a), while only small vertical scales are excited by the hyperbolic instability in pathway P2 (DNS2a). We now also see that this has a fundamental impact on total the viscous dissipation rate of the flow, which is substantially larger in P2 owing to the substantially smaller length scale of the vertical shear excited and dissipated.

By contrast, we see from figure\ref{fig:lengthscales}({\it d}) that the dissipation of buoyancy variance, and consequently the diapycnal mixing rate, is entirely controlled by flow on small vertical scales ($|k_z| \ge 10$) for both DNS1a and DNS2a. While this is expected for DNS2a (where only small scales are excited), the reason why this also applies for DNS1a is because the $k_z\neq 0$ modes of the primary instability satisfy the scaling $\hat b \sim Fr^2 k_z^2 \hat u \sim Fr^2 k_z^2  \hat v$ as long as $Fr k_z \ll 1$  \citep{Cocusseetal2025}. Thus, if $k_z^2 E(k_z)$ is roughly constant (and dominated by horizontal flows) we expect $k_z^2 B(k_z) \sim k_z^2 \hat b^2 \sim Fr^4 k_z^6 \hat u^2  \sim Fr^4 k_z^4$ at low wavenumbers, which is indeed the case (see dotted line in \ref{fig:lengthscales}({\it d})). This means that the dissipation of buoyancy variance is  negligible for $k_z \ll Fr^{-1}$ for DNS1a, and peaks when $k_z \sim Fr^{-1}$ as in DNS2a. The dissipation spectra of DNS1a and DNS2a therefore look quite similar, which explains why the buoyancy dissipation is more-or-less the same in the two simulations at the time of saturation. 

 These results therefore answer question (2) raised in \S\ref{sec:intro}: pathways P1 and P2 lead to a similar amount of diapycnal mixing but significantly different instantaneous mixing efficiencies in strongly stratified horizontal shear flows, primarily because they excite a very different spectrum of vertical wavenumbers, and consequently have different viscous dissipation rates. This statement is expected to hold whenever $Fr \ll 1$ and for sufficiently large $Re_b$, as long as the secondary instabilities in P2 only excite a narrow range of vertical scales around $L_b$. 
}

\section{Summary, Discussion and Conclusion}
\label{sec:ccl}

\pg{In this paper, we have studied  two possible pathways to the development of vertical shear starting from a vertically-invariant, horizontally-sheared, strongly stratified flow. This vertical shear can then excites small-scale KH instabilities and stratified turbulence when the buoyancy Reynolds number is sufficiently large. Both pathways can be achieved in the same model setup simply by changing the initial conditions slightly.

 In the first pathway (see P1 in figure \ref{fig:illustration}), initialized with white noise only, a broad spectrum of $k_z \ne 0$ modes develop through the primary horizontal shear instability of $\bar \bu$, as predicted by  \citet{Cocusseetal2025}}. These have large horizontal scales, and grow at a rate that is close  to that of the faster-growing $k_z = 0$ mode \pg{as long as $Frk_z \ll 1$}. \pg{They drive the growth of vertical shear, which later becomes unstable to secondary KH instabilities for sufficiently large $Re_b$. In the second pathway (P2 in figure \ref{fig:illustration}), the $k_z = 0$ mode is preferentially seeded in the initial conditions, and rapidly grows to dominate the flow dynamics. This causes a vertically-invariant} meandering of $\bar {\bf u}$, and then the formation of columnar vortices that are subject to 3D  secondary instabilities. \pg{In our Kolmogorov flow simulations, these secondary instabilities  have a high vertical wavenumber $k_z$, and are caused by a hyperbolic instability. The hyperbolic modes then drive the growth of vertical shear layers that eventually becomes unstable to tertiary vertical KH instabilities for sufficiently large $Re_b$. 

While these results were obtained specifically for the Kolmogorov flow, at $Pr = 1$, we believe that the coexistence of pathways P1 and P2 in strongly stratified, vertically-invariant horizontal shear flows is robust. Indeed, \citet{Cocusseetal2025} formally showed that as long as a flow of the form $\bar \bu = \bar u(y) \be_x$ is unstable to a $k_z = 0$ mode, a broad spectrum of $k_z \neq 0$ modes are also unstable in the limit of strong stratification, regardless of the flow shape and applied boundary conditions. This holds both at $Pr = O(1)$ and at $Pr \ll 1$ (with $Pe \ll 1$ while $Re \gg 1$). Thus, the existence of a $k_z = 0$ primary  mode necessarily opens up both pathways P1 and P2. Furthermore, the existence of several possible secondary instabilities of columnar flows in pathway P2, most of which give rise to the emergence of vertical structures on the buoyancy scale $L_b$, shows that P2 is robust as well. Thus, the P1 and P2 pathways can both be routes to turbulence at sufficiently large $Re_b = Fr^2 Re$.   

Which pathway is ultimately selected by the system depends on the ratio of the initial amplitude of the $k_z = 0$ mode relative to  the $k_z \neq 0$ modes for the specific initial conditions selected. We  hypothesize that the main reasons pathway P1 has not been clearly identified in simulations of a hyperbolic tangent layer to date could be because these simulations are often seeded with a $k_z = 0$ mode to speed up the computation, or, they have been run in a more weakly stratified limit where the range of unstable $k_z\neq 0$ modes of the primary instability is much more limited \citep{Deloncle2007,Cocusseal2026}. 

Interestingly, we have found that the peak dissipation of buoyancy variance $\chi$ is relatively independent of the pathway taken (for a given choice of $Fr, Re, Pe$). That is because this dissipation takes place primarily at small scales where $k_z \ge  Fr^{-1}$. By contrast, the 
peak viscous dissipation $\epsilon$ is significantly higher in P2 than in P1, owing to the intrinsically small vertical scale of the hyperbolic instability excited. As a result, we found that the mixing efficiency is significantly larger in pathway P1 than in P2. We believe this result should hold for any vertically-invariant horizontally-sheared flow with an inflection-point instability, as long as P1 leads to the excitation of a much broader range of vertical scales than P2, which is generally true when $Fr \ll 1$. For more weakly stratified flows, by contrast, where the range of excited scales is similar, and where most of the dissipation takes place in the KH billows rather than in the layered laminar flow around them, the two pathways should have similar mixing efficiencies.  
}

\pg{Our results} 
raise further questions that \pg{will} need to be answered in view of future applications to more realistic geophysical (and astrophysical) flows. First, we found that when $k_z = 0$ is dominant in the initial conditions, its growth appears to suppress the growth of the primary $k_z \neq 0$ modes of instability \pg{(even though they could in principle grow at a fairly similar rate)}, thus leading to the long-lived two-dimensional transient described in \S\ref{subsec:DNS2}. \pg{The two pathways thus appear to be mutually exclusive, at least for the idealized initial conditions applied to these 
simulations. This raises the interesting question of what pathway a flow  would take in less idealized, more natural conditions: on the one hand, the background flow is  more likely to be time-dependent, but at the same time, also less likely to have a dominant $k_z = 0$ seed. We also acknowledge the  possibility that neither pathways are particularly relevant, and that an ambient gravity wave field, or pre-existing vertical shear on small vertical scales \citep{LewinCaulfield2024}, are instead the key processes for the transition to turbulence in real systems.}

Second, we recognize that the effects of rotation must be taken into account in order to apply this work to large-scale, low Rossby number geophysical flows. \pg{Rotation could stabilize the primary instability of the flow $\bar \bu$, or its secondary instabilities.}
Based on the  work of \citet{HattoriHirota2023} on the stability of a rotating stratified Taylor-Green vortex array, we anticipate that rotation will stabilize some modes, and destabilize others. \pg{Progress on understanding the linear stability of an isolated hyperbolic point in both weakly and strongly  rotating flows, using WKB analysis, is under way \citep{RajkotiaZaheeretal2025}.

Overall, however, our work has demonstrated that a pre-existing vertical shear is not needed to trigger the onset of small-scale vertical KH instabilities in strongly stratified horizontally-sheared flows. Instead, the existence of not just one but at least two possible pathways for the excitation of $k_z \neq 0$ perturbations and the spontaneous emergence of small-scale vertical shear where none was initially present suggests that horizontal shear instabilities are likely key players in the generation and maintenance of stratified turbulence in geophysical and astrophysical flows. }

\begin{acknowledgements}
\pg{The authors thank Greg Chini, Colm-cille Caulfield, Sam Lewin, Marion Cocusse, and Chang Liu for their insight on stratified turbulence, and three anonymous referees for excellent suggestions that have helped improve the manuscript. } This work was initiated as a project for the Applied Mathematics 227 course offered in Winter 2024 at the University of California, Santa Cruz. P.G. and D.B. were later supported by NSF AST 2408025. \pg{Many of the later discussions were held at the 2025 GFD summer program held at the Woods Hole Oceanographic Institution. } The authors report no conflicts of interest. 
\end{acknowledgements}

\section*{Appendix A: the fastest-growing mode of instability of a sinusoidal jet.}
\label{sec:appA}

This appendix presents the linear stability analysis of $\bar {\bf u}$ (given by equation \ref{eq:ubardef}) to 2D and 3D triply-periodic perturbations, as discussed in \S \ref{sec:DNS}. \pg{This analysis effectively reproduces the work of \citet{Cocusseetal2025}, focussing on the case where the background flow is purely sinusoidal.}

\pg{We write the flow $\bu$ as
 \begin{equation}
     \bu = \bar \bu(y) + \tilde{\bu} (x,y,z,t) =\bar \bu(y) +  \hat \bu(y)  e^{\sigma t + i k_x x + i k_z z} ,
\label{eq:primaryansatz}
 \end{equation}
  (and similarly for the pressure and buoyancy perturbations $p$ and $b$) where $\sigma$ is the complex growth rate, $k_x$ is the streamwise wavenumber, and $k_z$ is the vertical wavenumber. Linearizing the governing equations around the background state,}
substituting the eigenmode ansatz (\ref{eq:primaryansatz}) into (\ref{eq:DNSumom})--(\ref{eq:DNSudiv}), and expanding it into components, we obtain:
\begin{eqnarray}
    \sigma \hat u + \hat v \cos(y) + i k_x \sin(y) \hat u = - i k_x \hat p + \frac{1}{Re}\left( \frac{d^2 \hat u}{d y^2} - (k_x^2 + k_z^2) \hat u \right), \label{eq:AppAu} \\
        \sigma \hat v + i k_x \sin(y)  \hat v = - \frac{d \hat p}{d y} + \frac{1}{Re} \left( \frac{d^2 \hat v}{d y^2} - (k_x^2 + k_z^2) \hat v \right),  \\
        \sigma \hat w + i k_x \sin(y)  \hat w = - i k_z \hat p+ \frac{\hat b}{Fr^2} + \frac{1}{Re} \left( \frac{d^2 \hat w}{d y^2} - (k_x^2 + k_z^2) \hat w \right),  \\
               \sigma \hat b + i k_x \sin(y)  \hat b + \hat w  =\frac{1}{Pe} \left( \frac{d^2 \hat b}{d y^2} - (k_x^2 + k_z^2) \hat b \right), \\ 
    i k_x \hat u + \frac{d \hat v}{d y} + i k_z \hat w = 0. \label{eq:AppAdiv}
\end{eqnarray}

We then assume that 
\begin{equation}
\hat q(y) = \sum_{n=-N}^N  q_n e^{i n y},
\label{eq:yexp}
\end{equation}
for $q \in \left\{ u,v,w,p,b\right\}$. Note that the dimensionless  wavenumber of the base flow is $k_y = 1$, so we omit it for simplicity of notation. This expression assumes that the perturbations have the same periodicity in $y$ as $\bar \bu$. \pg{Because the computational domain has dimension $L_y  = 2\pi$ in the spanwise direction, this is necessarily true for the DNS presented, but we note that modes with $k_y <1$  could also exist if the domain was not so restricted.}
 Also note that in theory the index $n$ should range from $-\infty$ to $+\infty$, but for numerical purposes the system of equations must be truncated to $n \in [-N,N]$. 

Then, using this Fourier decomposition   and projecting onto each mode yields a set of algebraic equations for the Fourier coefficients: 
\begin{eqnarray}
\sigma u_n + \frac{1}{2} (v_{n-1} + v_{n+1}) + \frac{k_x}{2}    (u_{n-1} - u_{n+1}) = - k_x \pi_n - \frac{k_x^2 + n^2+ k_z^2}{Re} u_n,  \\
        \sigma v_n + \frac{k_x}{2}    ( v_{n-1} -  v_{n+1})  = -  n \pi_n - \frac{k_x^2 + n^2 + k_z^2}{Re}  v_n,  \\
        \sigma w_n + \frac{k_x}{2}    ( w_{n-1} -  w_{n+1})  = -  k_z  \pi_n + \frac{ b_n}{Fr^2} - \frac{k_x^2 + n^2 + k_z^2}{Re}   w_n,  \\
        \sigma  b_n + \frac{k_x}{2}    ( b_{n-1} -  b_{n+1}) +  w_n =  - \frac{k_x^2 + n^2+ k_z^2}{Pe}   b_n,  \\
     k_x  u_n  + n  v_n +  k_z  w_n =0,
    \end{eqnarray}
    where we have defined $\pi_n \equiv i p_n$.  The system forms a  generalized  eigenvalue problem of the form $\mathbf{A} \mathbf{X} = \sigma \mathbf{B} \mathbf{X}$, where the matrices $\mathbf{A}$ and $\mathbf{B}$ are real, and where $\mathbf{X}$ is a vector containing the $ N_{\rm tot} = 5\times (2N+1)$ coefficients $\{ u_n\}$, $\{v_n\}$, $\{w_n\}$, $\{b_n\}$ and $\{\pi_n\}$. This type of problem can be solved using standard computational packages. Here, we use Matlab's {\bf eig} solver. For given values of the input parameters $Re$, $Pe$ and $Fr$, and each value of $k_x$, the solver returns $N_{\rm tot}$ eigenvalues $\sigma$, but we only retain the one with the largest real part (taking care to ignore singular eigenvalues), and call this the growth rate $\lambda(k_x,k_z)$.

    \begin{figure}
    \includegraphics[width=\textwidth]{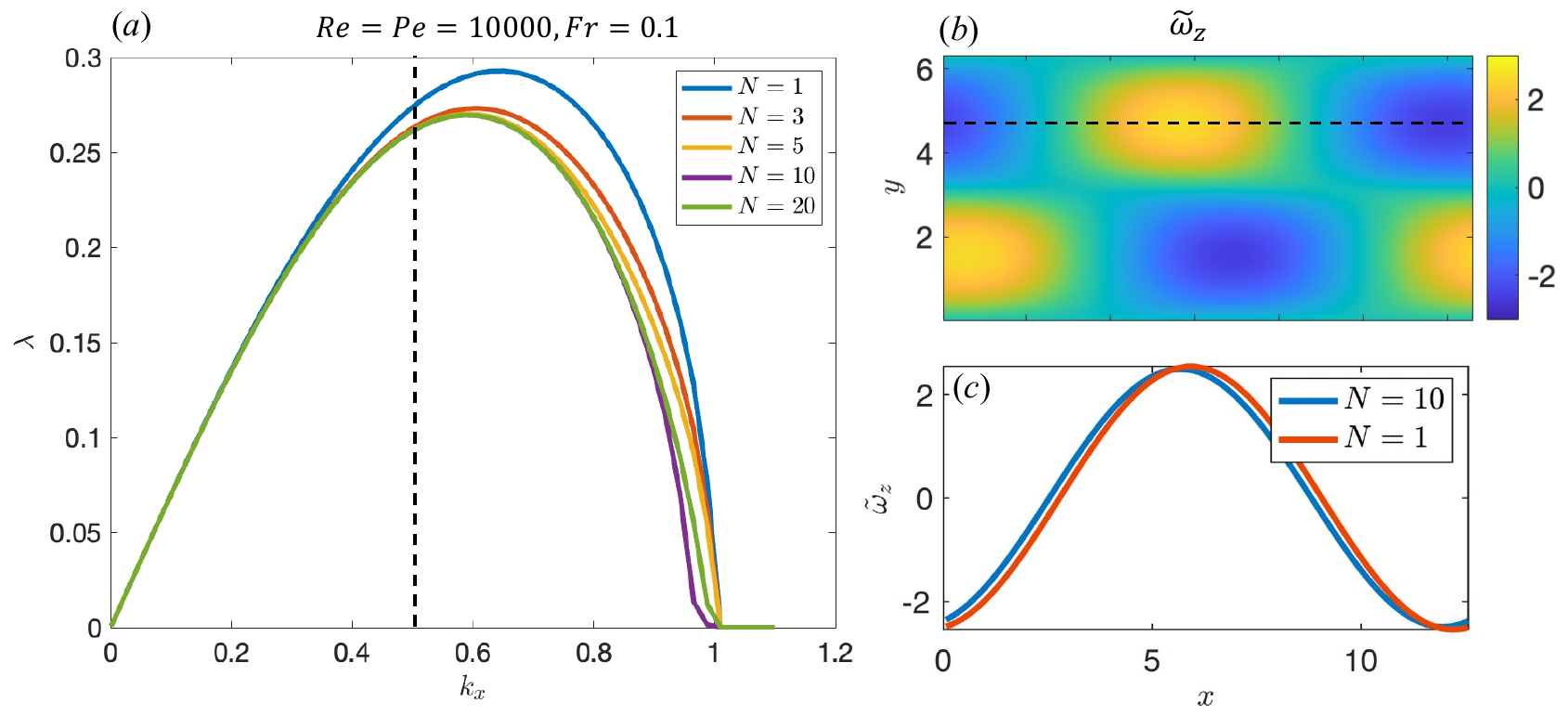}
    \caption{({\it a}) Real part of the growth rate, $\lambda$, as a function of $k_x$, for $Re = Pe = 10000$, $Fr = 0.1$, and $k_z = 0$, obtained using different Fourier resolutions (see legend). ({\it b}) Colormap of $\tilde{\omega}_z$ for the $k_x = 0.5$ mode, which is the fastest-growing mode in a domain of length $L_x = 4\pi$. The parameters are the same as in ({\it a}), with $N=10$. ({\it c}) Comparison of $\tilde{\omega}_z(x,3\pi/2)$, for the $N=1$ and $N=10$ resolutions. The parameters are otherwise the same as in ({\it a}).}
    \label{fig:vorticitymap}
    \end{figure}

While high-$k_z$ eigenmodes sometimes exhibit fine-scale horizontal structures that require a relatively large number of Fourier modes $N$ to be fully resolved, the $k_z = 0$ eigenmode can be resolved using a small $N$. This is illustrated in figure \ref{fig:vorticitymap}({\it a}), which shows the growth rate $\lambda = \Re(\sigma)$ for the $k_z = 0$ mode for $Re = Pe = 10000, Fr = 0.1$, keeping different number of Fourier modes. We see that relatively few modes are needed to recover an accurate $\lambda(k_x,0)$ curve, and that even the most dramatic truncation ($N=1$) provides a good qualitative fit to the higher-resolution simulation.

 We now fix the length of the domain to be $L_x = 4\pi$ (see main text for detail), with $k_x = 0.5$ and $k_z = 0$. The spatial structure of the corresponding mode can be obtained from the vector $\mathbf{X}$. Its associated vorticity field $\tilde \omega_z = \partial \tilde v /\partial x - \partial \tilde u/\partial y$ is shown in  
    figure \ref{fig:vorticitymap}({\it b}) using $N=10$ modes; the equivalent figure for the $N=1$ truncation looks almost identical. Small quantitative differences can be seen in figure \ref{fig:vorticitymap}({\it c}) which directly compares $\tilde \omega_z(x,y=3\pi/2)$ for the $N=1$ and $N=10$ truncations. We conclude that the $N=1$ truncation of fastest-growing, $k_z = 0$ mode is an acceptable approximation of the true linear eigenmode.

    The analytical expression for the $N=1$ truncation of $\hat \bu$ with $k_z = 0$ and $k_x = 0.5$ is henceforth denoted as $\hat \bu_{\rm m}$, and is equal to:
    \begin{eqnarray}
        \hat \bu_{\rm m}(y) = \left(e^{-iy} + e^{iy}\right)\be_x + \left( \frac{1}{2}e^{-iy} + v_0 - \frac{1}{2} e^{iy}\right)\be_y\nonumber \\ = 2 \cos(y) \be_x + \left( v_0 - i \sin(y) \right) \be_y.
    \end{eqnarray}
    where $v_0 \simeq - 0.913$ for large enough $Re$. The complete expression for this eigenmode is:  
    \begin{eqnarray} 
    \tilde{\bu}_{\rm m}(x,y) && = \Re\left[\hat \bu_{\rm m}(y) e^{i k_x x}\right]  \nonumber \\ 
&& =  2 \cos(y) \cos(k_x x) \be_x + \left( v_0 \cos(k_x x) + \sin(y) \sin(k_x x)  \right) \be_y .
\end{eqnarray}
It is easy to verify that $\nabla \cdot \tilde{\bu}_{\rm m} = 0$ because $k_x = 1/2$. 

\section*{Appendix B: Eigenvalue problem for the stability of the meandering flow}

This appendix presents the linear stability analysis of ${\bf u}_{\rm b}$ (given by equation \ref{eq:ub}) to two and three-dimensional triply-periodic perturbations, as discussed in \S \ref{sec:secondary}.
We begin with equations (\ref{eq:momsecond})--(\ref{eq:divsecond}) from the main text, and substitute the ansatz (\ref{eq:2Dansatz}), together with the expression for $\bu_{\rm b}$. We note that a recurring term in the momentum and buoyancy fluctuation equations is of the form $ \bu_{\rm b} \cdot \nabla_{\rm h} \hat q$. This term is expanded as
\begin{eqnarray}
\bu_{\rm b} \cdot \nabla_{\rm h} \hat q = u_{\rm b} \frac{\partial \hat q}{\partial x} + v_{\rm b} \frac{\partial \hat q}{\partial y} \nonumber  \\ 
= \frac{1}{2} \left[ -i \left( e^{i y} - e^{-iy} \right) + a \left( e^{i k_x x-iy} + e^{-ik_x x+iy} + e^{i k_x x+iy} + e^{-ik_x x-iy}   \right) \right] \nonumber  \\ \times \sum_{m,n}  i m k_x q_{mn} e^{i m k_x x + i ny}  \nonumber \\
+ \frac{a}{2} \left[ v_0\left( e^{i k_x x} +  e^{-i k_x x}\right) + \frac{1}{2} \left( e^{i k_x x - i y} + e^{- i k_x x + i y} -  e^{i k_x x + i y} -  e^{-i k_x x - i y}  \right) \right] \nonumber \\ \times \sum_{m,n}  i n  q_{mn} e^{i m k_x  x + i ny} \nonumber \\ 
= \frac{1}{2} \sum_{m,n}  m k_x q_{mn} \left( e^{i m k_x + i (n+1) y} -  e^{i m k_x + i (n-1) y} \right) \nonumber \\
+ \frac{ia}{2} \sum_{m,n}  m k_x q_{mn}  \left( e^{i (m+1) k_x x + i(n-1)y} + e^{i(m-1)k_x x+i(n+1)y} \right. \nonumber \\ \left. + e^{i (m+1)k_x x+i(n+1)y} + e^{i(m-1)k_x x+ i(n-1)y}   \right)    \nonumber \\
+ \frac{iav_0}{2} \sum_{m,n}   n  q_{mn} \left( e^{i (m+1) k_x x + i n y} +  e^{i(m-1) k_x x + i n y }\right) \nonumber \\ 
+ \frac{ia}{4} \sum_{m,n}  n  q_{mn} \left( e^{i (m+1) k_x x + i (n-1) y} + e^{i(m-1) k_x x + i (n+1) y} \right. \nonumber \\ \left. -  e^{i (m+1) k_x x + i (n+1) y} -  e^{i (m-1) k_x x + i(n-1) y}  \right).  
    \end{eqnarray}
Projecting this expression onto the Fourier modes yields
\begin{eqnarray}
   \frac{1}{L_xL_y} \int_x \int_y  (\bu_{\rm b} \cdot \nabla_{\rm h} \hat q)  e^{i m k_x x + i n y} dx dy  \nonumber \\ 
     = \frac{mk_x}{2} \left( q_{m,n-1} -q_{m,n+1} \right)    \nonumber  \\
     + \frac{i a k_x}{2}    \left( (m-1) ( q_{m-1,n+1} + q_{m-1,n-1}) + (m+1) (q_{m+1,n-1}   +  q_{m+1,n+1} ) \right)    
     \nonumber \\
+ \frac{ia nv_0}{2} \left(   q_{m-1,n} + q_{m+1,n} \right)  \nonumber \\ 
+ \frac{ia}{4} \left( (n-1) (q_{m+1,n-1} -  q_{m-1,n-1}) +  (n+1)  ( q_{m-1,n+1}  -  q_{m+1,n+1} ) \right)  .
    \end{eqnarray}

Similar steps can be used to expand and then project the terms $\hat \bu \cdot \nabla u_{\rm b}$ and $\hat \bu \cdot \nabla v_{\rm b}$, which result in 
\begin{eqnarray}
 \frac{1}{L_xL_y}\int\int (\hat \bu \cdot \nabla u_{\rm b})e^{i m k_x x + i n y} dxdy =  
 \frac{1}{2} (v_{m,n-1} + v_{m,n+1}) \nonumber  \\
+ \frac{ia k_x}{2}     \left( u_{m-1,n+1}  - u_{m+1,n-1} + u_{m-1,n-1} - u_{m+1,n+1} \right)   \nonumber \\ 
+ \frac{ia}{2} \left( - v_{m-1,n+1} + v_{m+1,n-1} + v_{m-1,n-1}  - v_{m+1,n+1} \right),
\end{eqnarray}
and 
\begin{eqnarray} 
\frac{1}{L_xL_y}\int\int (\hat \bu \cdot \nabla v_{\rm b})e^{i m k_x x + i n y} dxdy = 
 \frac{i a k_x v_0 }{2}  \left( u_{m-1,n} - u_{m+1,n} \right) \nonumber \\  + \frac{i a k_x }{4}  \left( u_{m-1,n+1}  - u_{m+1,n-1} - u_{m-1,n-1} + u_{m+1,n+1}  \right) \nonumber  \\
 + \frac{ia}{4} \left( - v_{m-1,n+1}  + v_{m+1,n-1} - v_{m-1,n-1}  + v_{m+1,n+1}   \right).
\end{eqnarray}
  
With the projection of these terms computed, the remaining equations are easy to obtain. We have:
\begin{itemize}
\item {the $x$-component of the momentum equation:}
\begin{eqnarray}
\sigma u_{mn} + \frac{mk_x}{2} \left( u_{m,n-1} -u_{m,n+1} \right)+ 
 \frac{1}{2} (v_{m,n-1} + v_{m,n+1})  
 \nonumber \\
+ \frac{ia}{2} \left( - v_{m-1,n+1} + v_{m+1,n-1} + v_{m-1,n-1}  - v_{m+1,n+1} \right)  
     \nonumber  \\
     + \frac{i a m k_x}{2}    \left( u_{m-1,n+1} + u_{m-1,n-1} + u_{m+1,n-1}   +  u_{m+1,n+1}  \right)    
    \nonumber  \\
+ \frac{ia nv_0}{2} \left(   u_{m-1,n} + u_{m+1,n} \right)  \nonumber  \\ 
+ \frac{ia}{4} \left( (n-1) (u_{m+1,n-1} -  u_{m-1,n-1}) +  (n+1)  ( u_{m-1,n+1}  -  u_{m+1,n+1} ) \right) \nonumber  \\
= - i m k_x p_{mn} - \frac{(mk_x)^2 + n^2 + k_z^2}{Re} u_{mn},
\label{eq:appmommn}
\end{eqnarray}
\item {the $y$-component of the momentum equation:}
\begin{eqnarray}
    \sigma v_{mn} + \frac{mk_x}{2} \left( v_{m,n-1} -v_{m,n+1} \right) + \frac{i a k_x v_0 }{2}  \left( u_{m-1,n} - u_{m+1,n} \right) \nonumber \\  + \frac{i a k_x }{4}  \left( u_{m-1,n+1}  - u_{m+1,n-1} - u_{m-1,n-1} + u_{m+1,n+1}  \right) \nonumber\\
     + \frac{i a k_x}{2}    \left( (m-1) ( v_{m-1,n+1} + v_{m-1,n-1}) + (m+1) (v_{m+1,n-1}   +  v_{m+1,n+1} ) \right)    
     \nonumber\\
+ \frac{ia nv_0}{2} \left(   v_{m-1,n} + v_{m+1,n} \right)  \nonumber\\ 
+ \frac{ina}{4} \left(  v_{m+1,n-1} -  v_{m-1,n-1} +   v_{m-1,n+1}  -  v_{m+1,n+1}  \right)\nonumber \\ 
 = - inp_{mn} - \frac{(mk_x)^2 + n^2 + k_z^2}{Re} v_{mn},
\end{eqnarray}
\item {the $z$-component of the momentum equation}
\begin{eqnarray}
    \sigma w_{mn} +  \frac{mk_x}{2} \left( w_{m,n-1} -w_{m,n+1} \right)  \nonumber    \\
     + \frac{i a k_x}{2}    \left( (m-1) ( w_{m-1,n+1} + w_{m-1,n-1}) + (m+1) (w_{m+1,n-1}   +  w_{m+1,n+1} ) \right)    
     \nonumber  \\
+ \frac{ia nv_0}{2} \left(   w_{m-1,n} + w_{m+1,n} \right)  \nonumber  \\ 
+ \frac{ia}{4} \left( (n-1) (w_{m+1,n-1} -  w_{m-1,n-1}) +  (n+1)  ( w_{m-1,n+1}  -  w_{m+1,n+1} ) \right) \nonumber  \\ = - ik_zp_{mn} + \frac{b_{mn}}{Fr^2} - \frac{(mk_x)^2 + n^2 + k_z^2}{Re} w_{mn},
\end{eqnarray}
\item {the buoyancy equation}:
\begin{eqnarray} 
\sigma b_{mn}  + \frac{mk_x}{2} \left( b_{m,n-1} -b_{m,n+1} \right)   \nonumber   \\
     + \frac{i a k_x}{2}    \left( (m-1) ( b_{m-1,n+1} + b_{m-1,n-1}) + (m+1) (b_{m+1,n-1}   +  b_{m+1,n+1} ) \right)    
     \nonumber  \\
+ \frac{ia nv_0}{2} \left(   b_{m-1,n} + b_{m+1,n} \right)  \nonumber  \\ 
+ \frac{ia}{4} \left( (n-1) (b_{m+1,n-1} -  b_{m-1,n-1}) +  (n+1)  ( b_{m-1,n+1}  -  b_{m+1,n+1} ) \right) + w_{mn} 
 \nonumber \\ = - \frac{(mk_x)^2 + n^2 + k_z^2}{Pe} b_{mn},
\end{eqnarray}
\item {the continuity equation:}
\begin{equation}
    mk_x u_{mn} + n v_{mn} + k_z w_{mn} = 0.
    \label{eq:appcontmn}
\end{equation}
\end{itemize}
It easy to check that when the Fourier expansion in $x$ only contains the $m=1$ term, and when $a = 0$ (i.e. with no meander), we recover the primary instability equations from Appendix A. Equations (\ref{eq:appmommn})--(\ref{eq:appcontmn}) form a generalized  eigenvalue problem of the type ${\bf AX} = \sigma {\bf BX}$, where ${\bf X}$  is a vector of length $N_{\rm tot} = 5 (2M+1) (2N+1)$, that can be solved numerically using Matlab's {\bf eig} solver. 

When $M$ and $N$ are large, the computational cost of solving this eigenvalue problem rapidly becomes prohibitive. In that case we use (\ref{eq:appcontmn}) to write $w_{mn}$ explicitly in terms of $u_{mn}$ and $v_{mn}$, namely
\begin{equation}
    w_{mn} = - k_z^{-1} ( mk_x u_{mn} + n v_{mn}),
\end{equation}
and substitute it into the remaining equations to reduce the size of the problem to $N_{\rm tot} = 4(2M+1)(2N+1)$. Note that this option is only possible when $k_z \neq 0$.
 
 For given values of the input parameters $Re$, $Pe$, $Fr$, $a$ and each value of $k_z$ (noting that $k_x= 1/2$ is fixed here, see main text for detail), the solver returns $N_{\rm tot}$ eigenvalues $\sigma$. Again, we only retain the one with the largest real part (taking care to ignore singular eigenvalues), and call this the growth rate $\lambda(k_z)$. In the main text, we are generally interested in the instability eigenmodes associated with the maximum of the $\lambda(k_z)$ curve and must therefore ensure that a sufficient number of Fourier modes $M$ and $N$ are kept to obtain an accurate estimate of their growth rate and eigenmode structure.

 Figure \ref{fig:compareMN} compares the growth rates obtained for 
 a given set of input parameters ($Re = Pe = 10000$, $Fr = 0.1$ and $a = 1$),
 using different  values of the Fourier resolution $M$. In each case, we set $N = M/2$ so the implied spatial resolution is the same in the $x$ and $y$ directions (i.e. $Mk_x = Nk_y)$, remembering that $k_x = 0.5$ and $k_y = 1$. The left panel shows the $\lambda(k_z)$ curves for different values of $M$. Note that the lowest value of $k_z$ used is $k_z = 0.01$ (not $k_z = 0$). We see that for low $k_z$ a relatively low Fourier resolution  is sufficient to estimate $\lambda$ with good accuracy. For larger values of $k_z$, however, a correspondingly larger value of $M$ is required for $\lambda$ to converge. The right panel shows this more quantitatively, by focusing on three different values of $k_z$. We see that, roughly speaking, $M$ needs to satisfy $Mk_x = Nk_y \ge k_z$ (denoted by the corresponding vertical dashed lines) to ensure that $\lambda$ has converged within a few percents of its true value (shaded box). This is consistent with the fact that the hyperbolic instability at high $k_z$ contains small horizontal scales. 
 
 In the main text, we are primarily interested in identifying the value of $k_z$ which maximizes the growth rate. The peak usually lies at moderate values of $k_z$ for the parameters explored, and so we have found that using $M = 20, N=10$ as a baseline, and checking that convergence has been achieved by comparing the results with the lower resolution $M = 16, N=8$ (and discarding cases for which the results do not match), is sufficient. 
  
\begin{figure}
    \includegraphics[width=\textwidth]{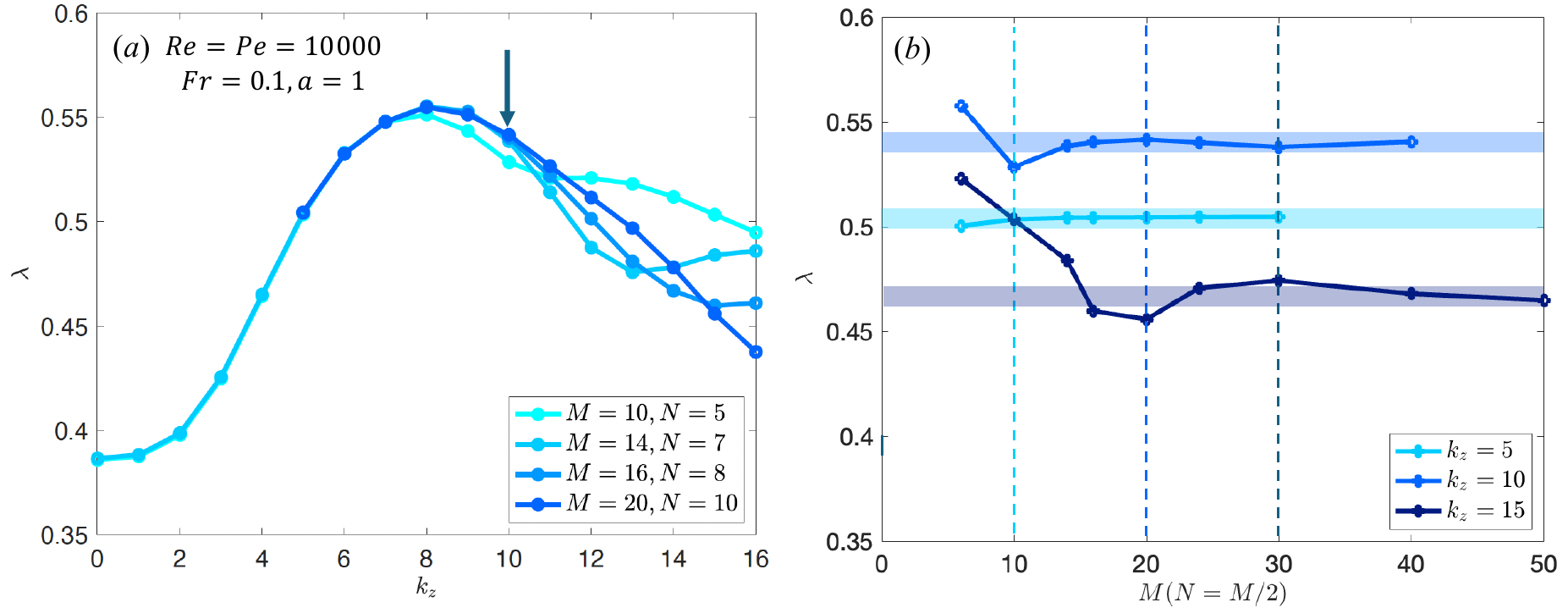}
    \caption{Analysis of the effect of the selected Fourier resolution $M$ (with $N=M/2)$ on the mode growth rate, for $Re = Pe = 10000$, $k_x = 0.5$, $Fr = 0.1$ and $a = 1$. ({\it a}) Comparison of $\lambda(k_z)$ obtained with different resolutions,  shown in the legend. The black arrow shows the largest value of $k_z$ for which we estimate that the growth rate $\lambda$ is accurately computed using $M=20$, $N=10$, by comparison with $M=16$, $N=8$. ({\it b}) Estimated growth rate as a function of $M$ for three different values of $k_z$. The vertical dashed lines show the value of $M$ for which $Mk_x = k_z$, and the shaded horizontal bars have a thickness of  $0.01$, which is approximately equal to two percent of $\lambda$.}
    \label{fig:compareMN}
    \end{figure}

The eigenmode structure is obtained by reconstructing the flow field $\tilde{\bu}$ from the coefficients $\{u_{mn},v_{mn},w_{mn}\}$. We have found that substantially higher values of $M$ and $N$ are needed for this purpose. Figure \ref{fig:moderes} compares the vertical velocity field $w'$ and the vertical vorticity field $\omega'_z = \partial_x v' - \partial_y u'$ obtained for two different computations of the fastest-growing eigenmode at parameters $Re = Pe = 10000$, $Fr = 0.1$, $a = 1$, which has $k_z = 8$. The lower resolution computation has $M = 20, N=10$, and the higher resolution computation has $M=40$, $N=20$. 
Even though the growth rates computed are almost the same in both cases (and equal to $\lambda \simeq 0.553$), we see that the low resolution computation is unable to capture the fine structure of this eigenmode. Even with $M = 40$, residual small-amplitude oscillations due to the Gibbs phenomenon remain, but at least the structure of the mode is clearly visible. In the main text, we use $M=50$ and $N=25$, which is the highest resolution we were able to achieve with the available computational tools at our disposition. The corresponding structure of the same eigenmode with that resolution is shown in figure \ref{fig:Modecompare}.

\begin{figure}
    \includegraphics[width=\textwidth]{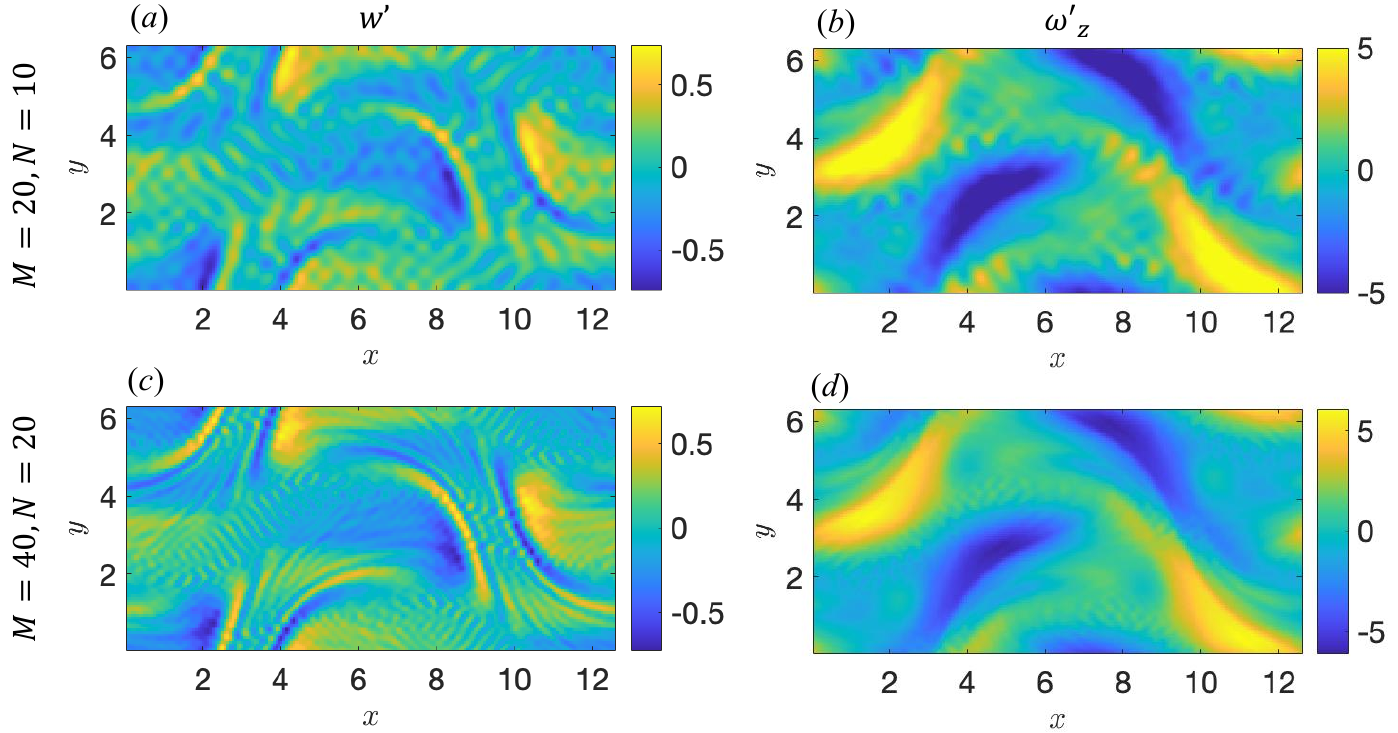}
    \caption{Visualizations of the vertical velocity $w'$ (panels ({\it a}) and ({\it c})) and of the vertical vorticity $\omega'_z$ (panels ({\it b}) and ({\it d})) of the fastest-growing mode of instability of  ${\bu}_{\rm b}$ for $Re = Pe = 10000$, $Fr = 0.1$, $a = 1$ and $k_z = 8$. The top row (panels ({\it a}) and ({\it b})) shows a 'low' resolution solution with $M =20$, $N=10$ and the bottom row (panels ({\it c}) and ({\it d}))  shows a higher-resolution one with $M = 40$, $N=20$. Note that both $w'$ and $\omega'_z$ vary sinusoidally in $z$, so the flow is shown in a representative horizontal plane at $z = \pi/(4k_z)$. }
    \label{fig:moderes}
    \end{figure}

\bibliographystyle{jfm}
\bibliography{References}

\end{document}